\documentclass[10 pt]{article}

\usepackage{amsmath,epsfig,amsfonts,graphicx,
amsgen,amsbsy,amsopn,amstext,amssymb}

\begin{document}

\title{Bose-Einstein Condensates in Superlattices}

\author{Mason A. Porter \\ mason@caltech.edu \\ Department of Physics and Center for the Physics of Information \\ California Institute of Technology, Pasadena, CA  91125, USA \\ \\
P. G. Kevrekidis \\ Department of Mathematics and Statistics \\ University of 
Massachusetts, Amherst MA  01003, USA}

\maketitle

\begin{abstract}
  
  We consider the Gross-Pitaevskii (GP) equation in the presence of
  periodic and quasiperiodic superlattices to study cigar-shaped
  Bose-Einstein condensates (BECs) in such potentials.  We examine
  spatially extended wavefunctions in the form of modulated amplitude
  waves (MAWs). With a coherent structure ansatz, we derive amplitude
  equations describing the evolution of spatially modulated states of
  the BEC.  We then apply second-order multiple scale perturbation
  theory to study harmonic resonances with respect to a single lattice
  wavenumber as well as ultrasubharmonic resonances that result from
  interactions of both wavenumbers of the superlattice.  In each case,
  we determine the resulting system's equilibria, which represent
  spatially periodic solutions, and subsequently examine the stability
  of the corresponding solutions by direct simulations of the GP
  equation, identifying them as typically stable solutions of the
  model.  We then study subharmonic resonances using Hamiltonian
  perturbation theory, tracing robust, spatio-temporally periodic
  patterns.

\end{abstract}

\subsection*{PACS: 05.45.-a, 03.75.Lm, 05.30.Jp, 05.45.Ac}
\subsection*{MSC: 37N20, 35Q55, 81V45}
\subsection*{Keywords: Bose-Einstein condensates, multiple-scale perturbation theory, Hamiltonian systems}
\maketitle

\vspace{2mm}

\section{Introduction}

At very low temperatures, particles in a dilute bose gas can occupy the same
quantum (ground) state, forming a Bose-Einstein condensate (BEC)
\cite{pethick,stringari,ketter,edwards}, which appears as a sharp peak (over a broader distribution) in both coordinate and momentum space. 
%(see Fig.~\ref{3peak}).  
As the gas is cooled, condensation (of a large fraction of the atoms in the gas) 
occurs via a quantum 
phase transition, emerging when the wavelengths of individual atoms overlap 
and behave identically.  Atoms of mass $m$ and temperature $T$ constitute 
quantum wavepackets whose spatial extent is given by the de Broglie wavelength 
\begin{equation}
	\lambda_{db} = \sqrt{\frac{2\pi \hbar^2}{mk_BT}}\,,
\end{equation}	
which represents the uncertainty in position associated with the momentum distribution \cite{ketter} (where $\hbar$ is Planck's constant and $k_B$ is Boltzmann's constant).  The atomic wavepackets overlap once atoms are cooled sufficiently so that $\lambda_{db}$ is comparable to the separation between atoms, as bosonic atoms then undergo a quantum phase transition to form a BEC (a coherent cloud of atoms).  Although a condensate constitutes a quantum phenomenon, such \textquotedblleft matter waves" can often be observed macroscopically, with the number of condensed atoms $N$ ranging from several thousand (or less) to several million (or more) \cite{stringari}.  

BECs were first observed experimentally in 1995 in dilute alkali gases such
as vapors of rubidium and sodium
\cite{becrub,becna}.  In these experiments, atoms were confined in
magnetic traps, evaporatively cooled to a fraction of a microkelvin,
left to expand by switching off the confining trap, and subsequently
imaged with optical methods.  A sharp peak in the velocity
distribution was observed below a critical temperature, 
indicating that condensation had occured [as the alkali atoms were now condensed
in the same (ground)
state].   Under the typical confining conditions of experimental settings, BECs are inhomogeneous, so condensates arise as a narrow peak not only in momentum space but also in coordinate space.  
% In uniform gases, by contrast, particles condense into a state with zero momentum.  (That notwithstanding, the onset of condensation is revealed using momentum-space observations, as condensed and non-condensed particles occupy the same volume in coordinate space.) \cite{stringari}  

%\begin{figure}[tbp]
%\centerline{
%\epsfig{file=3peaks.eps, width=14cm,angle=0, clip=}
%\caption{Formation of a Bose-Einstein condensate, as shown in the 1995 MIT experiments.  The colors give the number of atoms at a given velocity (with blue indicating fewer atoms and red indicating more atoms).  A peak forms (that is, condensation occurs) as the temperature $T$ goes below the critical temperature $T_c$.  (Image courtesy W. Ketterle's BEC group, MIT.)} 
%\label{3peak}
%\end{figure}

The macroscopic observability of the condensates in coordinate and momentum space 
has led to novel methods of investigating quantities such as energy and density distributions, 
interference phenomena, the frequencies of collective excitations, the temperature dependence of BECs, among others \cite{stringari} (for comprehensive reviews, the interested reader
should consult Refs.~\cite{pethick,string}).  Another consequence of this inhomogeneity is that the effects of two-body interactions are greatly enhanced, despite the fact that bose gases are 
extremely dilute (with the average distance between atoms typically more than 
ten times the range of interatomic forces).  For example, these interactions 
reduce the condensate's central density and enlarge the size of the condensate 
cloud, which becomes macroscopic and can be measured directly with optical 
imaging methods.

BECs have two characteristic length scales.  The condensate density varies on the scale of the harmonic
oscillator length $a_{ho} = \sqrt{\hbar/(m\omega_{ho})}$ [which is typically on
the order of a few microns], where $\omega_{ho}=(\omega_x \omega_y
\omega_z)^{1/3}$ is the geometric mean of the trapping frequencies.  The \textquotedblleft coherence length" (or \textquotedblleft healing length"), determined by balancing the quantum pressure and the condensate's interaction energy, is $\chi=1/\sqrt{8\pi |a| \bar{n}}$ [and is also typically of the order of a few microns],
%can range from about half a micron to about 20 microns], 
where $\bar{n}$ is the mean particle density and $a$, the (two-body) $s$-wave scattering length, is determined by the atomic species of the condensate.  Interactions between atoms are repulsive when $a > 0$ and attractive when $a < 0$.  For a dilute ideal gas, $a \approx 0$. The length scales in BECs should be contrasted with those in systems like
superfluid helium, in which the effects of inhomogeneity occur on a
microscopic scale fixed by the interatomic distance \cite{stringari}.

If considering only two-body, mean-field interactions, a dilute
Bose-Einstein gas near zero temperature can be modeled using a cubic nonlinear Schr\"odinger
equation (NLS) with an external potential, which is also known as the
Gross-Pitaevskii (GP) equation.  This is written \cite{stringari}
\begin{equation}
  	i\hbar {\Psi }_{t}=\left( -\frac{\hbar ^{2}\nabla ^{2}}{2m}+g_{0}|\Psi|^{2}+{\mathcal{V}}(\vec{r})\right) \Psi\,,  \label{GPE}
\end{equation}
where $\Psi =\Psi (\vec{r},t)$ is the condensate wave function
normalized to the number of atoms, ${\mathcal{V}}(\vec{r})$ is the external potential, and the
effective interaction constant is $g_{0}=[4\pi\hbar ^{2}a/m][1+O(\zeta ^{2})]$, where $\zeta \equiv
\sqrt{|\Psi |^{2}|a|^{3}}$ is the dilute-gas parameter \cite{stringari,kohler,baiz}. 

BECs are modeled in the quasi-one-dimensional (quasi-1D) regime when the transverse dimensions
of the condensate are on the order of its healing length and its
longitudinal dimension is much larger than its transverse ones
\cite{bronski,bronskirep,bronskiatt,stringari}.  In this regime, one
employs the 1D limit of a 3D mean-field theory (generated by averaging
in the transverse plane) rather than a true 1D mean-field theory,
which would be appropriate were the transverse dimension on the order
of the atomic interaction length or the atomic size \cite{bronski,salasnich,towers}.  The resulting 1D equation is \cite{salasnich,stringari}
\begin{equation}
	      i\hbar u_t = -[\hbar^2/(2m)] u_{xx} + g|u|^2 u + V(x) u \,, \label{nls3}
\end{equation}
where $u$, $g$, and $V$ are, respectively, the rescaled 1D wave 
function (\textquotedblleft order parameter"), interaction constant, and external trapping potential.  The quantity $|u|^2$ gives the atomic number density.  The self-interaction parameter $g$ is tunable 
(even its sign), because the scattering length $a$ can be adjusted using 
magnetic fields in the vicinity of a Feshbach resonance \cite{fesh,FRM}.  The manipulation of Feshbach resonances has become one of the most active areas in the study of ultracold atoms, as (for example) numerous research groups are investigating the intermediate regime between molecular condensates and degenerate Fermi gases (the so-called \textquotedblleft BEC-BCS" crossover regime).  Theoretical algorithms for manipulating $a$, such as alternating it periodically between positive and negative values, have been developed by analogy with \textquotedblleft dispersion management" in nonlinear optics.

In forming a BEC, the atoms are trapped using a confining magnetic or optical potential $V(x)$, which is then turned off so that the gas can expand and be imaged.  In early experiments, only parabolic (\textquotedblleft harmonic") potentials were employed, but a wide variety of potentials can now be constructed experimentally.  In addition to harmonic traps, these include 
double-well traps (see, e.g., \cite{markus} and references therein), 
periodic lattices (see, e.g., \cite{konotopmplb} for a review), and 
superlattices \cite{quasibec,anamaria} (which can be either periodic or
quasiperiodic), and superpositions of lattices or superlattices with
harmonic traps.  Optical lattices and superlattices are created using counter-propagating laser beams, and higher-dimensional versions of many of the aforementioned potentials have also been achieved experimentally.  

The existence of quasi-1D (``cigar-shaped'') BECs motivates the study of lower dimensional models such as Eq. (\ref{nls3}).  The case of periodic and quasiperiodic potentials
without a confining trap along the longitudinal dimension of the lattice is of
particular theoretical and experimental interest.  Such potentials
have been used, for example, to study Josephson effects
\cite{anderson}, squeezed states \cite{squeeze}, Landau-Zener
tunneling and Bloch oscillations \cite{morsch}, and the transition
between superfluidity and Mott insulation at both the classical
\cite{smerzi,cata} and quantum \cite{mott} levels.  Moreover, with
each lattice site occupied by one alkali atom in its ground state,
BECs in optical lattices show promise as a register in a quantum
computer \cite{porto,voll}.

In experiments, a weak harmonic trap is typically used on top of the
optical lattice (OL) or optical superlattice (OSL) to prevent the
particles from escaping.  The lattice is also generally turned on
after the trap.  If one wishes to include the trap in theoretical
analyses, then $V(x)$ is modeled by
\begin{equation}
        V(x) = V_1\cos(\kappa_1 x) + V_2\cos(\kappa_2 x) + V_h x^2 \,, 
\label{harmwiggle}
\end{equation}
where $\kappa_1$ is the primary lattice wavenumber, $\kappa_2 >
\kappa_1$ is the secondary lattice wavenumber, $V_1$ and $V_2$ are the
associated lattice amplitudes, and $V_h$ represents the magnitude of
the harmonic trap.  Note that $V_1$, $V_2$, $V_h$, $\kappa_1$, and
$\kappa_2$ can all be tuned experimentally, so that the external potential's length scales are easily manipulated.  The sinusoidal terms in (\ref{harmwiggle}) dominate for small $x$, but the harmonic trap otherwise becomes quickly dominant.  When $V_h \ll V_1\,,V_2$, the potential is
dominated by its periodic (or quasiperiodic) contributions for many periods
\cite{promislow,mapbec}.  BECs in OLs with up to $200$
wells have been created experimentally \cite{well}.  

In this work, we let $V_h = 0$ and focus on OL and OSL potentials.
Spatially periodic potentials have been employed in experimental
studies of BECs \cite{hagley,anderson,squeeze,morsch,mott,porto} and
have also been studied theoretically
\cite{bronski,space1,space2,malopt,alf,pethick2,machholm,mueller,mapbecprl,smerzi,band,njppgk}; see also the recent reviews \cite{pk1,pk2}.
In experiments reported in 2003, BECs were loaded into OSLs with $\kappa_2 =
3\kappa_1$ \cite{quasibec}.  However, there has thus far been very
little theoretical research on BECs in superlattice potentials
\cite{anamaria,mottsuper,louis2,eksioglu}.  In this work, we consider
both periodic (rational $\kappa_2/\kappa_1$) and quasiperiodic
(irrational $\kappa_2/\kappa_1$) OSLs.
 
We focus here on spatially extended solutions rather than on localized waves (solitons).  For BECs loaded into OSLs, the interest in such extended wavefunctions is twofold.  First, BECs were successfully loaded into OSL potentials in recent experiments \cite{quasibec} (in which extended solutions were observed).  Second, MAWs in BECs in OSLs can be used to study period-multiplied states and generalizations thereof \cite{mapbecprl,mapbec,mapbin}.

On the first front, $^{87}$Rb atoms were loaded into an OSL by the sequential creation of two
lattice structures.  The atoms were initially loaded into every third
site of an OL.  A second periodic structure was subsequently added so
that the atoms could be transferred from long-period lattice sites to
corresponding short-period lattice sites in a patterned loading.  

On the second front, Machholm {\it et al.} \cite{pethick2} studied
period-doubled states (in $|u|^2$), interpreting them as
soliton trains to attempt to explain experimental studies by Cataliotti{\it et al.} \cite{cata}, who
observed superfluid current disruption in chains of weakly coupled
BECs in OL potentials.  More recently, experimental observations of period doubled wavefunctions in BECs in OL potentials have now been reported \cite{chu}.  From a dynamical systems perspective,
period-multiplied states arise at the center of KAM islands in phase
space; the location and size of these islands have been estimated using
Hamiltonian perturbation theory and multiple scale analysis
\cite{mapbecprl,mapbec,mapbin}.

In this study, we investigate spatially extended solutions of BECs in
periodic and quasiperiodic OSLs.  We apply a coherent structure ansatz
to Eq. (\ref{nls3}), yielding a parametrically forced Duffing
equation describing the spatial evolution of the field.  We employ
second-order multiple scale perturbation theory to study its periodic
orbits (called \textquotedblleft modulated amplitude waves'' and
denoted MAWs), and illustrate their dynamical stability with numerical
simulations of the GP equation.  We consider harmonic ($1\!:\!1$)
resonances and two types of ultrasubharmonic resonances---resulting
from, respectively, ``additive'' ($2\!:\!1+1$) and ``subtractive''
($2\!:\!1-1$) interactions---all of which arise at the
$O(\varepsilon^2)$ level of analysis.  Because ultrasubharmonic
resonances result from the interaction of multiple superlattice
wavenumbers, they cannot occur in BECs loaded into regular OLs.  We
then explore subharmonic resonances using Hamiltonian perturbation
theory, identifying various relevant patterns including
quasi-stationary ones (with weak amplitude oscillations) and
spatio-temporally breathing ones (see the details below).
 
We structure the rest of our presentation as follows: We first introduce modulated
amplitude waves and use multiple scale perturbation theory to derive
``slow flow'' dynamical equations that describe the resonance
phenomena under consideration. We analyze these equations and
corroborate our results and test the stability of the
MAWs with direct numerical simulations of the GP equation.  We then
examine subharmonic resonances using Hamiltonian perturbation theory
and additional numerics.  Finally, we summarize our findings and
present our conclusions.

\section{Modulated Amplitude Waves} \label{mawsec}

To study MAWs, we employ the ansatz
\begin{equation}
        u(x,t) = R(x)\exp\left(i\left[\theta(x) - \mu t\right]\right) 
\,.  \label{maw2}
\end{equation}
When these (temporally periodic) coherent structures (\ref{maw2})
are also spatially periodic, they are called {\it modulated amplitude
  waves} (MAWs) \cite{lutz1,lutz2}.  The orbital stability of MAWs for
the cubic NLS with elliptic potentials has been studied by Bronski {\it et al} \cite{bronski,bronskiatt,bronskirep}.  To obtain stability
information about sinusoidal potentials, one takes the limit as the
elliptic modulus $k$ approaches zero \cite{lawden}.  When $V(x)$ is
periodic, the resulting MAWs generalize the Bloch modes that occur in
the theory of linear systems with periodic potentials
\cite{675,ashcroft,band,space1,space2}.  In this work, we extend
recent studies \cite{mapbecprl,mapbec} of the dynamical behavior of
MAWs of BECs in lattice potentials to superlattice potentials.

Inserting Eq. (\ref{maw2}) into Eq. (\ref{nls3}), 
equating the real and imaginary components of the resulting equation, 
and defining $S := R'$ yields
%\begin{align}
%       \hbar\mu R(x) &= -\frac{\hbar^2}{2m}R''(x) + \left[\frac{\hbar^2}{2m}\left[\theta'(x)\right]^2 + gR^2(x) + V(x) \right]R(x) \,, \\
%       0 &= \frac{\hbar^2}{2m}\left[2\theta'(x)R'(x) + \theta''(x)R(x)\right] 
%            \,, \notag
%\end{align}
%which results in 
the following two-dimensional system of nonlinear 
ordinary differential equations:
\begin{align}
        R' &= S \,, \notag \\
        S' &= \frac{c^2}{R^3} - \frac{2m\mu R}{\hbar} 
+ \frac{2mg}{\hbar^2}R^3 + \frac{2m}{\hbar^2}V(x)R \,. \notag
\end{align}
The parameter $c$ is given by the relation
\begin{equation}
        \theta'(x) = \frac{c}{R^2} \,,   \label{angmom}
\end{equation}
which indicates conservation of ``angular momentum'' \cite{bronski}.
Constant phase solutions (i.e., standing waves), which constitute an
important special case, satisfy $c = 0$.  In the rest of the paper, we
restrict ourselves to this class of solutions, so that
\begin{align}
        R' &= S \,, \notag \\
        S' &= - \frac{2m\mu R}{\hbar} 
+ \frac{2mg}{\hbar^2}R^3 + \frac{2m}{\hbar^2}V(x)R \,. \label{dynam35}
\end{align}

We consider the case with $V_h = 0$ (which implies, in practice, that
the harmonic trap is negligible with respect to the OSL potential for
the domain of interest) and define
\begin{equation}
        \tilde{\delta} := \frac{2m\mu}{\hbar} \,, \quad
        \varepsilon\tilde{\alpha} := -\frac{2mg}{\hbar^2} \,, \quad 
        \tilde{V}(x) := -\frac{2m}{\hbar^2}V(x)\,,
\end{equation}
where
\begin{equation}
        \tilde{V}(x) = \varepsilon[\tilde{V}_1\cos(\kappa_1x) 
+ \tilde{V}_2\cos(\kappa_2 x)] \,,
\end{equation}
the parameters $\tilde{\delta}$, $\tilde{\alpha}$, and $\tilde{V}_j$
are $O(1)$ quantities, and the lattice wavenumbers $\kappa_j$ can either be commensurate
(rational multiples of each other) or incommensurate, so that the OSL
can be, respectively, either periodic or quasiperiodic.  We let
$\kappa_2 > \kappa_1$ without loss of generality, so that $\kappa_1$
is the primary lattice wavenumber.  In our numerical
simulations, we focus on the case $\kappa_2=3 \kappa_1$ which has been
achieved experimentally \cite{quasibec}.

For notational convenience, we drop the tildes from $\tilde{\delta}$, 
$\tilde{\alpha}$, and $\tilde{V}_j$, so that Eq. (\ref{dynam35}) is 
written
\begin{equation}
        R'' + \delta R + \varepsilon\alpha R^3 
+ \varepsilon R[V_1\cos(\kappa_1 x) + V_2\cos(\kappa_2 x)] = 0 \,. \label{sup}
\end{equation}
In this paper, we consider the case $\delta > 0$ corresponding to a 
positive chemical potential.

\section{Multiple Scale Perturbation Theory and Spatial Resonances} 
\label{multiple}

To employ multiple scale perturbation theory \cite{bo,675}, we
define ``slow space'' $\eta := \varepsilon x$ and ``stretched space''
\begin{equation}
        \xi := bx = [1 + \varepsilon b_1 + \varepsilon^2 b_2 
+ O(\varepsilon^3)]x \,.
\label{eqb}
\end{equation}
We then expand the wavefunction amplitude $R$ in a power series,
\begin{equation}
        R = R_0 + \varepsilon R_1 + \varepsilon^2 R_2 + O(\varepsilon^3)\,, 
\label{eqr}
\end{equation}
and stretch the spatial dependence in the OSL potential, which is then
written
\begin{equation}
        \bar{V}(\xi) = V_1\cos(\kappa_1 \xi) + V_2\cos(\kappa_2 \xi)\,.
\end{equation}
Inserting these expansions, Eq. (\ref{sup}) becomes
\begin{align}
        &\left[1 + b_1\varepsilon + b_2\varepsilon^2 + O(\varepsilon^3)\right]^2\left[\frac{\partial^2 R_0}{\partial \xi^2} + \varepsilon\frac{\partial^2 R_1}{\partial \xi^2} + \varepsilon^2\frac{\partial^2 R_2}{\partial \xi^2} + O(\varepsilon^3)\right] \notag \\ &\qquad + 2\varepsilon\left[1 + b_1\varepsilon + b_2\varepsilon^2 + O(\varepsilon^3)\right]\left[\frac{\partial^2 R_0}{\partial \xi \partial \eta} + \varepsilon\frac{\partial^2 R_1}{\partial \xi \partial \eta} + \varepsilon^2\frac{\partial^2 R_2}{\partial \xi \partial \eta} + O(\varepsilon^3)\right] \notag \\ &\qquad + \varepsilon^2\left[\frac{\partial^2 R_0}{\partial \eta^2} + \varepsilon\frac{\partial^2 R_1}{\partial \eta^2} + \varepsilon^2\frac{\partial^2 R_2}{\partial \eta^2} + O(\varepsilon^3)\right] \notag \\ &\qquad + \delta\left[R_0 + \varepsilon R_1 + \varepsilon^2 R_2 + O(\varepsilon^3)\right] + \varepsilon\alpha\left[R_0 + \varepsilon R_1 + \varepsilon^2 R_2 + O(\varepsilon^3)\right]^3 \notag \\ &\qquad + \varepsilon\left[R_0 + \varepsilon R_1 + \varepsilon^2 R_2 + O(\varepsilon^3)\right]\left[V_1\cos(\kappa_1 \xi) + V_2\cos(\kappa_2 \xi)\right] = 0\,.
\end{align}

To perform multiple scale analysis, we equate the coefficients of terms of 
different order (in $\varepsilon$) in turn.  At $O(1) = O(\varepsilon^0)$, we
 obtain
\begin{equation}
        \frac{\partial^2 R_0}{\partial \xi^2} + \delta R_0 = 0\,, \notag
\end{equation}
which has the solution
\begin{equation}
        R_0(\xi,\eta) = A(\eta)\cos(\sqrt{\delta}\xi) 
+ B(\eta)\sin(\sqrt{\delta}\xi)\,, \label{ro}
\end{equation}
for slowly-varying amplitudes $A(\eta)$, $B(\eta)$, equations of motion 
for which arise at $O(\varepsilon)$.

Equating coefficients at $O(\varepsilon)$ yields
\begin{align}
        \frac{\partial^2 R_1}{\partial \xi^2} + \delta R_1 &=
\left[2b_1\delta A - 2\sqrt{\delta}B' - \frac{3}{4}\alpha A(A^2 + B^2)\right]\cos(\sqrt{\delta}\xi) 
\notag \\&\quad + \left[2b_1\delta B + 2\sqrt{\delta}A' - \frac{3}{4}\alpha B(A^2 + B^2)\right]\sin(\sqrt{\delta}\xi) \notag \\ &\quad + \frac{\alpha A}{4}[-A^2 + 3B^2]\cos(3\sqrt{\delta}\xi) + \frac{\alpha B}{4}[-3A^2 + B^2]\sin(3\sqrt{\delta}\xi) \notag \\ &\quad + \frac{V_1A}{2}\cos([\kappa_1 - \sqrt{\delta}]\xi) + \frac{V_1A}{2}\cos([\kappa_1 + \sqrt{\delta}]\xi) \notag \\ &\quad - \frac{V_1B}{2}\sin([\kappa_1 - \sqrt{\delta}]\xi) + \frac{V_1B}{2}\sin([\kappa_1 + \sqrt{\delta}]\xi) \notag \\ &\quad + \frac{V_2A}{2}\cos([\kappa_2 - \sqrt{\delta}]\xi) + \frac{V_2A}{2}\cos([\kappa_2 + \sqrt{\delta}]\xi) \notag \\ &\quad - \frac{V_2B}{2}\sin([\kappa_2 - \sqrt{\delta}]\xi) + \frac{V_2B}{2}\sin([\kappa_2 + \sqrt{\delta}]\xi)\,. \label{o1}
\end{align}

For $R_1(\xi,\eta)$ to be bounded, the coefficients of the secular
terms in Eq. (\ref{o1}) must vanish \cite{675,bo}.  The harmonics
$\cos(\sqrt{\delta}\xi)$ and $\sin(\sqrt{\delta}\xi)$ are always
secular, whereas $\cos(3\sqrt{\delta}\xi)$ and
$\sin(3\sqrt{\delta}\xi)$ are never secular.  The other harmonics are
secular only in the case of $2\!:\!1$ subharmonic resonances
\cite{mapbecprl,mapbec}, which can occur with respect to either the
primary ($\kappa_1 = 2\sqrt{\delta}$) or secondary ($\kappa_2 =
2\sqrt{\delta}$) wavenumber of the lattice.  We will consider the
situation in which (\ref{o1}) is non-resonant and turn our attention
to other resonant situations at $O(\varepsilon^2)$ that arise from
interactions between both lattice wavenumbers.  [Our
$O(\varepsilon^2)$ analysis below can be repeated in the presence of
$2\!:\!1$ resonances.]  At $O(\varepsilon)$, one obtains either
no resonance, a long-wavelength subharmonic resonance, or a
short-wavelength subharmonic resonance.

Equating the coefficients of the secular terms to zero in Eq. (\ref{o1}) 
yields the following equations of motion describing the slow dynamics:
\begin{align}
        A' &= -b_1\sqrt{\delta}B + \frac{3\alpha}{8\sqrt{\delta}}B(A^2 + B^2) 
\,, \notag \\
        B' &= b_1\sqrt{\delta}A - \frac{3\alpha}{8\sqrt{\delta}}A(A^2 + B^2) 
\,.   \label{slow1}
\end{align}
We convert (\ref{slow1}) to polar coordinates with $A(\eta) = C
\cos[\varphi(\eta)]$ and $B(\eta) = C \sin[\varphi(\eta)]$ and see immediately that each circle of constant $C$ is invariant.  The dynamics on each circle is given by 
\begin{equation}
  \varphi(\eta) = \varphi(0) + \left[b_1\sqrt{\delta} 
- \frac{3\alpha}{8\sqrt{\delta}}C^2\right]\eta\,.
\end{equation}
We examine the special circle of equilibria, corresponding to periodic orbits of (\ref{nls3}), which satisfies
\begin{equation}
        C^2 = A^2 + B^2 = \frac{8b_1\delta}{3\alpha}\,. \label{echo}
\end{equation}

We are interested in the $O(\varepsilon^2)$ effects, which we now
analyze.  At this second order of perturbation theory, BECs in OSL
potentials exhibit dynamical behavior that cannot occur in BECs in
simpler OL potentials (where, for example, solutions of type of Eq.
(\ref{echo}) straightforwardly arise \cite{mapbin}).

Equating coefficients at $O(\varepsilon^2)$ yields
\begin{align}
        \frac{\partial^2 R_2}{\partial \xi^2} + \delta R_2 &= -(b_1^2 + 2b_2)\frac{\partial^2R_0}{\partial \xi^2} - \frac{\partial^2R_0}{\partial \eta^2} - 2b_1\frac{\partial^2 R_0}{\partial \xi \partial \eta} - 3\alpha R_0^2R_1 - 2b_1\frac{\partial^2 R_1}{\partial \xi^2} - 2\frac{\partial^2 R_1}{\partial \xi \partial \eta} \notag \\ &\quad - R_1V_1\cos(\kappa_1\xi) - R_2V_2\cos(\kappa_2\xi)\,, \label{2eq}
\end{align}
where one inserts the expressions for $R_0$, $R_1$, and their derivatives into
 the right-hand-side of (\ref{2eq}).

To find the secular terms in Eq. (\ref{2eq}), we compute
\begin{align}
        R_1(\xi,\eta) &= C(\eta)\cos(\sqrt{\delta}\xi) + D(\eta)\sin(\sqrt{\delta}\xi) + R_{1p}(\xi,\eta)\,, \notag \\
        R_{1p}(\xi,\eta) &= c_1\cos(3\sqrt{\delta}\xi) + c_2\sin(3\sqrt{\delta}\xi) \notag \\ &\quad + \sum_{j = 1}^2 \left[c_3^j\cos([\kappa_j - \sqrt{\delta}]\xi) + c_4^j\cos([\kappa_j + \sqrt{\delta}]\xi) + c_5^j\sin([\kappa_j - \sqrt{\delta}]\xi) + c_6^j\sin([\kappa_j + \sqrt{\delta}]\xi)   \right]\,, \label{r1}
\end{align}
where $j \in \{1\,,2\}$ and 
\begin{align}
        c_1 &= \frac{\alpha}{32\delta}A(A^2 - 3B^2)\,, \quad
        c_2 = \frac{\alpha}{32\delta}B(3A^2 - B^2)\,, \notag \\
        c_3^j &= \frac{V_jA}{2\kappa_j(\kappa_j-2\sqrt{\delta})}\,, \quad
        c_4^j = \frac{V_jA}{2\kappa_j(\kappa_j+2\sqrt{\delta})}\,, \notag \\
        c_5^j &= \frac{V_jB}{2\kappa_j(\kappa_j - 2\sqrt{\delta})}\,, \quad
        c_6^j = \frac{V_jB}{2\kappa_j(\kappa_j+2\sqrt{\delta})}\,.
\end{align}

Inserting (\ref{ro}) and (\ref{r1}) into Eq. (\ref{2eq}) and expanding
the resulting equation trigonometrically yields $19$ harmonics (that
are also present for sines), which we list in Table \ref{harm}.  We
indicate which of these harmonics are always secular, sometimes
secular, and never secular.

At this order of perturbation theory, one finds $2\!:\!1$ (primary
subharmonic), $4\!:\!1$ (secondary subharmonic), $1\!:\!1$ (harmonic),
$2\!:\!1\!+\!1$ (additive ultrasubharmonic), and $2\!:\!1\!-\!1$
(subtractive ultrasubharmonic) resonances.  The first three types of
resonances can occur with respect to either $\kappa_1$ or $\kappa_2$,
whereas the latter two require the interaction of both superlattice
wavenumbers.  Harmonic and ultrasubharmonic spatial resonances have
not been analyzed previously for BECs, and subharmonic resonances have
only been analyzed in the case of regular OL potentials.  At
$O(\varepsilon)$, we considered the case without $2\!:\!1$ resonances,
so the associated resonance conditions ($\kappa_j = \pm
2\sqrt{\delta}$) are necessarily not satisfied at the present
[$O(\varepsilon^2)$] stage, as indicated in Table \ref{harm}.
Second-order subharmonic ($4\!:\!1$) resonances have been studied in
BECs in regular OL potentials \cite{mapbecprl,mapbec}.  Their associated resonance
conditions are $\kappa_j = \pm 4\sqrt{\delta}$.  (We return to
subharmonic resonances in the case of OSLs later when we apply
Hamiltonian perturbation theory.)   The resonance
relations for harmonic resonances are $\kappa_j = \pm \sqrt{\delta}$.
We will consider solutions that have harmonic resonance with respect
to the primary lattice wavenumber $\kappa_1$.  The resonance relation
for additive ultrasubharmonic resonances is $\kappa_2 + \kappa_1 = \pm
2\sqrt{\delta}$, and that for subtractive ultrasubharmonic resonances
is $\kappa_2 - \kappa_1 = \pm 2\sqrt{\delta}$.  In the remainder of
this section, we consider the non-resonant, harmonically resonant, and
the two types of ultrasubharmonic resonant states in turn.

It is also important to remark that with the slow spatial variable
$\eta = \varepsilon x$, the approximate solutions $R(x)$ obtained
perturbatively are valid for $|x| \lesssim O(\varepsilon^{-1})$
despite the fact that we employ a second-order multiple scale
expansion.  By incorporating a third (``super slow'') scale
$\varepsilon^2 x$, which is more technically demanding, one can obtain
approximate solutions that are valid for $|x| \lesssim
O(\varepsilon^{-2})$ \cite{bo}.

\begin{table}[t] 
\centerline{
\begin{tabular}{|l|c|c|r|} \hline
Label & Harmonic & Secular? & Resonance when secular  \\ \hline
1 & $\cos(\sqrt{\delta}\xi)$ & Yes   &  N/A   \\
2 & $\cos(3\sqrt{\delta}\xi)$ & No   &  N/A  \\
3 & $\cos(5\sqrt{\delta}\xi)$ & No   &  N/A  \\
4 & $\cos([\kappa_1 - \sqrt{\delta}]\xi)$ & Assumed not in resonance at $O(\varepsilon)$   &  $2\!:\!1$  \\
5 & $\cos([\kappa_1 + \sqrt{\delta}]\xi)$ & Assumed not in resonance at $O(\varepsilon)$  & $2\!:\!1$   \\
6 & $\cos([\kappa_2 - \sqrt{\delta}]\xi)$ & Assumed not in resonance at $O(\varepsilon)$  & $2\!:\!1$   \\
7 & $\cos([\kappa_2 + \sqrt{\delta}]\xi)$ &  Assumed not in resonance at $O(\varepsilon)$  &  $2\!:\!1$ \\
8 & $\cos([\kappa_1 - 3\sqrt{\delta}]\xi)$ & Sometimes  &  $4\!:\!1$    \\
9 & $\cos([\kappa_1 + 3\sqrt{\delta}]\xi)$ & Sometimes  &  $4\!:\!1$    \\
10 & $\cos([\kappa_2 - 3\sqrt{\delta}]\xi)$ & Sometimes & $4\!:\!1$     \\
11 & $\cos([\kappa_2 + 3\sqrt{\delta}]\xi)$ & Sometimes & $4\!:\!1$     \\
12 & $\cos([2\kappa_1 - \sqrt{\delta}]\xi)$ & Sometimes & $1\!:\!1$     \\
13 & $\cos([2\kappa_1 + \sqrt{\delta}]\xi)$ & Sometimes & $1\!:\!1$     \\
14 & $\cos([2\kappa_2 - \sqrt{\delta}]\xi)$ & Sometimes & $1\!:\!1$     \\
15 & $\cos([2\kappa_2 + \sqrt{\delta}]\xi)$ & Sometimes & $1\!:\!1$     \\
16 & $\cos([\kappa_1 + \kappa_2 - \sqrt{\delta}]\xi)$ & Sometimes & $2\!:\!1\!+\!1$ \\
17 & $\cos([\kappa_1 + \kappa_2 + \sqrt{\delta}]\xi)$ & Sometimes & $2\!:\!1\!+\!1$ \\
18 & $\cos([\kappa_1 - \kappa_2 - \sqrt{\delta}]\xi)$ & Sometimes & $2\!:\!1\!-\!1$ \\
19 & $\cos([\kappa_1 - \kappa_2 + \sqrt{\delta}]\xi)$ & Sometimes & $2\!:\!1\!-\!1$  \\ \hline
\end{tabular}}\caption{The harmonics in the right-hand-side of 
Eq. (\ref{2eq}) after the formulas for $R_0$ (\ref{ro}) and $R_1$ (\ref{r1}) 
are inserted.  We only list the cosines in this table, but the sines of these
 harmonics are present as well.  We designate which harmonics are always 
secular, sometimes secular (under an appropriate resonance condition, as 
detailed in the text), and never secular.}\label{harm}
\end{table}

Before proceeding, we also remark that in light of KAM theory, one
expects different dynamical behavior (at least mathematically)
depending on whether $\kappa_2/\kappa_1$ is an integer, a rational
number, or an irrational number.  Only the situation $\kappa_2 =
3\kappa_1$ has been prepared experimentally, so we concentrate on that
case in our numerical simulations.  

We note additionally that we simulated the dynamics and examined the
stability of MAWs using a numerical domain with periodic boundary
conditions. This allows us to handle integer or rational values of
$\kappa_2/\kappa_1$ with appropriate selection of the domain
parameters (so that the box size is an integer multiple of both
spatial periods). However, quasiperiodic potentials cannot be tackled
numerically within this framework for the extended wave solutions
considered in this section.  Our analytical work on MAWs is valid for
all real ratios $\kappa_2/\kappa_1$.

\subsection{The Non-Resonant Case}

In the non-resonant case, effective equations governing the 
$O(\varepsilon^2)$ slow evolution are 
\begin{align}
        C' &= \frac{1}{\Delta(\delta,\kappa_1,\kappa2)}\left[\left( f_1(\alpha,\delta,\kappa_1,\kappa_2)B^2 + f_2(\alpha,\delta,\kappa_1,\kappa_2)A^2 + f_3(\alpha,\delta,\kappa_1,\kappa_2,b_1) \right)D + f_4(\alpha,\delta,\kappa_1,\kappa_2)ABC \right. \notag \\ &\qquad \left.+ f_5(\alpha,\delta,\kappa_1,\kappa_2)B^5 + f_6(\alpha,\delta,\kappa_1,\kappa_2)A^2B^3 + f_7(\alpha,\delta,\kappa_1,\kappa_2)A^4B + f_8(\alpha,\delta,\kappa_1,\kappa_2,b_2)B \right] \,, \notag \\
        D' &= -\frac{1}{\Delta(\delta,\kappa_1,\kappa2)}\left[\left( f_1(\alpha,\delta,\kappa_1,\kappa_2)A^2 + f_2(\alpha,\delta,\kappa_1,\kappa_2)B^2 + f_3(\alpha,\delta,\kappa_1,\kappa_2,b_1) \right)C + f_4(\alpha,\delta,\kappa_1,\kappa_2)ABD \right. \notag \\ &\qquad \left. + f_5(\alpha,\delta,\kappa_1,\kappa_2)A^5 + f_6(\alpha,\delta,\kappa_1,\kappa_2)A^3B^2 + f_7(\alpha,\delta,\kappa_1,\kappa_2)AB^4 + f_8(\alpha,\delta,\kappa_1,\kappa_2)A \right] \,, \label{nonres}
\end{align}
where
\begin{equation}
        \Delta(\delta,\kappa_1,\kappa_2) = 256\delta^{3/2}\left(16\delta^2 - 4\delta\kappa_1^2 -4\delta\kappa_2^2 + \kappa_1^2\kappa_2^2\right)
\end{equation}
and
\begin{align}
        f_1(\alpha,\delta,\kappa_1,\kappa_2) &= 3 f_2(\alpha,\delta,\kappa_1,\kappa_2)   \,, \notag \\
        f_2(\alpha,\delta,\kappa_1,\kappa_2) &= 96\alpha\delta[16 \delta^2 - 4\delta(\kappa_1^2 + \kappa_2^2) + \kappa_1^2\kappa_2^2]  \,, \notag \\
        f_3(\alpha,\delta,\kappa_1,\kappa_2,b_1) &= 256\delta^2b_1[-\kappa_1^2\kappa_2^2 + 4\delta(\kappa_1^2 + \kappa_2^2) -16\delta^2]  \,, \notag \\
        f_4(\alpha,\delta,\kappa_1,\kappa_2) &= 2 f_2(\alpha,\delta,\kappa_1,\kappa_2)  \,, \notag \\
        f_5(\alpha,\delta,\kappa_1,\kappa_2) &= 15\alpha^2[-16\delta^2 + 4\delta(\kappa_1^2 + \kappa_2^2) - \kappa_1^2\kappa_2^2]  \,, \notag \\
        f_6(\alpha,\delta,\kappa_1,\kappa_2) &= 2 f_5(\alpha,\delta,\kappa_1,\kappa_2)  \,, \notag \\
        f_7(\alpha,\delta,\kappa_1,\kappa_2) &= f_5(\alpha,\delta,\kappa_1,\kappa_2)  \,, \notag \\
        f_8(\alpha,\delta,\kappa_1,\kappa_2,b_2) &= 64\delta[V_1^2\kappa_2^2 + V_2^2\kappa_1^2 -4\delta(V_1^2 + V_2^2 + \kappa_1^2\kappa_2^2b_2) + 16\delta^2b_2(\kappa_1^2 + \kappa_2^2) - 64\delta^3b_2]   \,.  \label{nono}
\end{align}
In this case, the OSL does not contribute to $O(\varepsilon^2)$ terms.

Equilibrium solutions  of (\ref{nonres}) satisfy 
\begin{align}
        C &= \frac{(f_1B^2 + f_2A^2 + f_3)(f_5A^5 + f_6A^3B^2 + f_7AB^4 + f_8A) - (f_4AB)(f_5B^5 + f_6A^2B^3 + f_7A^4B + f_8B)}{f_4^2A^2B^2 -(f_1B^2 + f_2A^2 + f_3)(f_1A^2 + f_2B^2 + f_3)} \,, \notag \\
        D &= \frac{(f_1A^2 + f_2B^2 + f_3)(f_5B^5 + f_6A^2B^3 + f_7A^4B + f_8B) - (f_4AB)(f_5A^5 + f_6A^3B^2 + f_7AB^4 + f_8A)}{f_4^2A^2B^2 -(f_1B^2 + f_2A^2 + f_3)(f_1A^2 + f_2B^2 + f_3)}\,, \label{order}
\end{align}
where one inserts an equilibrium value of $A$ and $B$ from Eq.
(\ref{echo}).  One then inserts equilibrium values of $A$, $B$, $C$,
and $D$ into Eqs. (\ref{ro}) and (\ref{r1}) to obtain the spatial
profile $R = R_0 + \varepsilon R_1 + O(\varepsilon^2)$ used as the
initial wavefunction in the numerical simulations of the full GP given
by Eq. (\ref{nls3}).

\begin{figure}[tbp]
%\centerline{
\epsfig{file=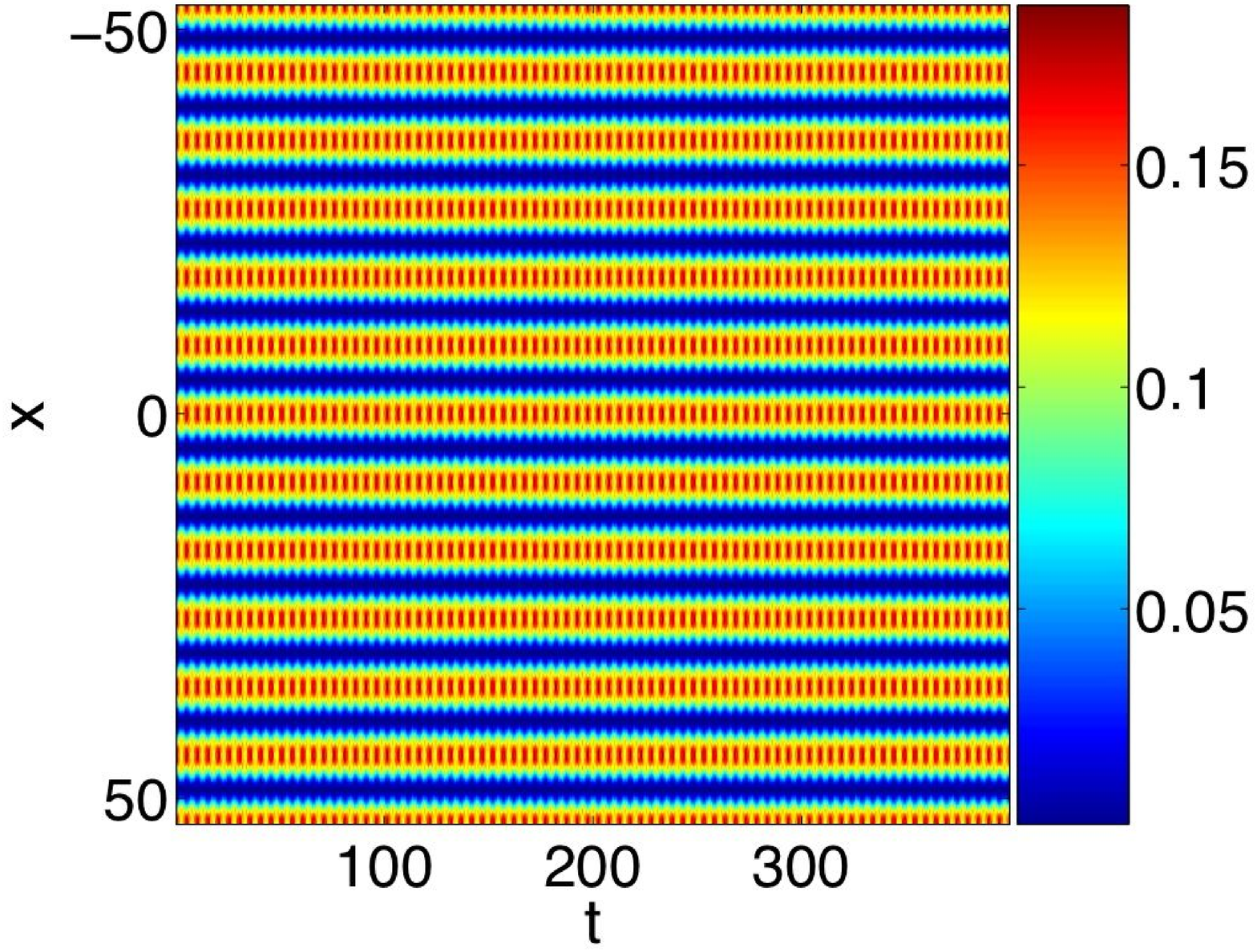, width=7.5cm,angle=0, clip=}
\epsfig{file=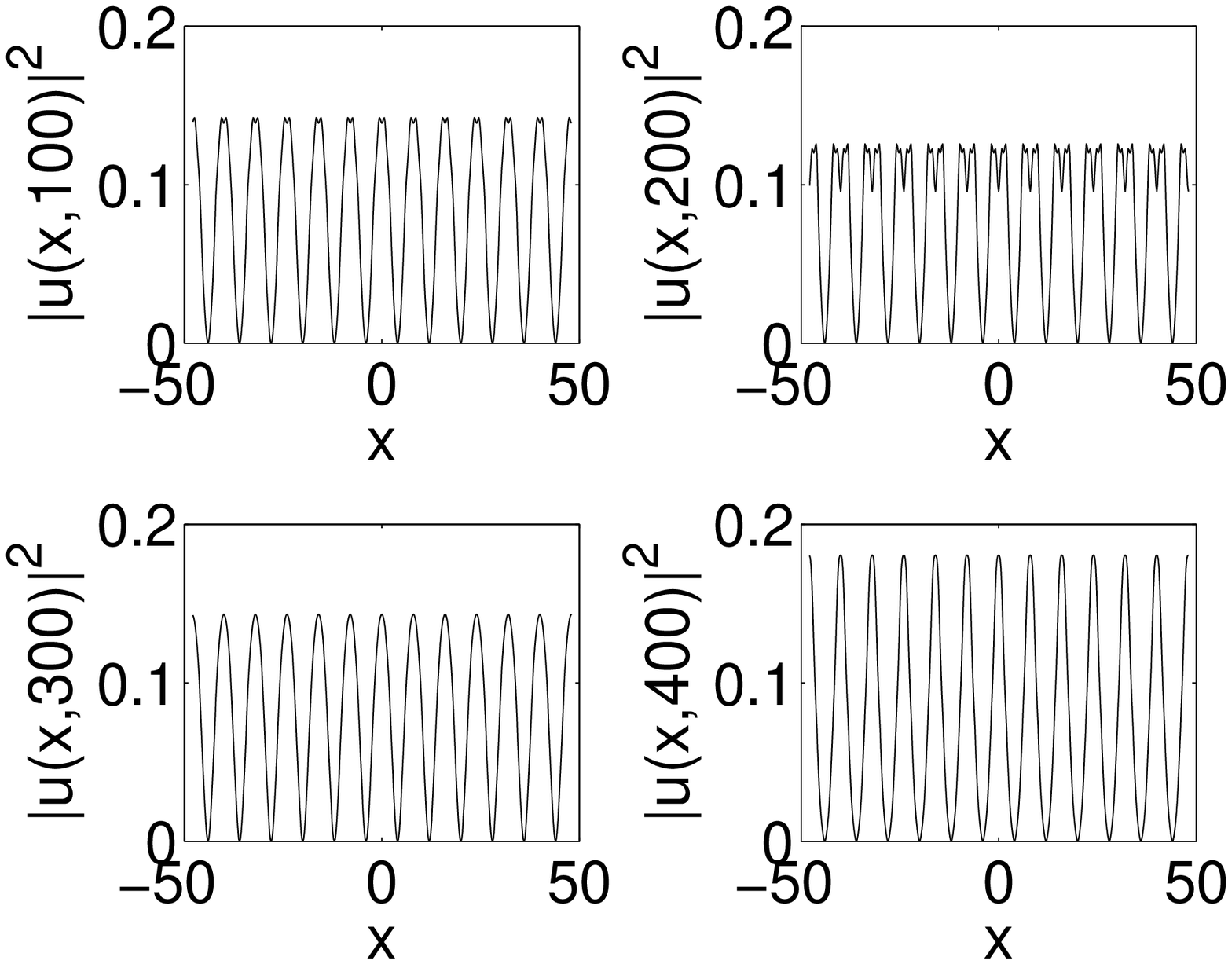, width=7.5cm,angle=0, clip=}
\caption{Evolution of the non-resonant spatially extended solution of Eq. 
  (\ref{eqr}) with $C$ and $D$ in Eq. (\ref{r1}) given by Eq.
  (\ref{nonres}) [see text for parameter details] for an OSL potential
  with $V_2=2 V_1=2$ and $\kappa_2=3 \kappa_1=3$.  The left panel
  shows the spatio-temporal evolution of $|u(x,t)|^2$ by means of
  a colored contour plot.  The right panel shows spatial profiles of
  $|u|^2$ at four values of time ($t=100$, $200$, $300$, and $400$).} 
\label{mpsuper4}
\end{figure}

A typical example of the non-resonant case is shown in Fig. \ref{mpsuper4}, 
with $V_2=2 V_1=2$ and $\kappa_2=3 \kappa_1=12 \sqrt{\delta}=3 \pi/(2 b)$, 
where $b$ is the stretching factor given by Eq. (\ref{eqb}).  In this 
simulation, we used $b_1=b_2=1$ and $\epsilon=0.1$. It can be clearly seen 
that the relevant solution is dynamically stable, which we found to be 
generic in our numerical experiments.  Simulations with rational 
$\kappa_2/\kappa_1$ reveal similar phenomena.

\subsection{Resonances}

In this subsection, we consider harmonic resonances, additive
ultrasubharmonic resonances, and subtractive ultrasubharmonic
resonances.  In the evolution equations for the slow dynamics, one
inserts the appropriate resonance relation into $\Delta$ and
$f_1$--$f_7$.  The function $f_8$ has both the non-resonant
contributions discussed above and additional resonant terms due to the
OSL.  Note additionally that there is a symmetry-breaking in
the resulting equations because the functional form of the lattice
contains only cosine terms.

\subsubsection{Harmonic Resonances}

When $\kappa_j = \pm \sqrt{\delta}$, there is a harmonic resonance.  The 
effective equations governing the $O(\varepsilon^2)$ slow evolution in the 
presence of a harmonic resonance with respect to the primary lattice wave 
number $\kappa_1$ are
\begin{align}
        C' &= \frac{1}{\Delta(\kappa_1,\kappa2)}\left[\left( f_1(\alpha,\kappa_1,\kappa_2)B^2 + f_2(\alpha,\kappa_1,\kappa_2)A^2 + f_3(\alpha,\kappa_1,\kappa_2,b_1) \right)D + f_4(\alpha,\kappa_1,\kappa_2)ABC \right. \notag \\ &\quad \left.+ f_5(\alpha,\kappa_1,\kappa_2)B^5 + f_6(\alpha,\kappa_1,\kappa_2)A^2B^3 + f_7(\alpha,\kappa_1,\kappa_2)A^4B + f_{8s}(\alpha,\kappa_1,\kappa_2,b_2)B \right] \,, \notag \\
        D' &= \frac{1}{\Delta(\kappa_1,\kappa2)}\left[\left( f_1(\alpha,\kappa_1,\kappa_2)A^2 + f_2(\alpha,\kappa_1,\kappa_2)B^2 + f_3(\alpha,\kappa_1,\kappa_2,b_1) \right)C + f_4(\alpha,\kappa_1,\kappa_2)ABD \right. \notag \\ &\quad \left. + f_5(\alpha,\kappa_1,\kappa_2)A^5 + f_6(\alpha,\kappa_1,\kappa_2)A^3B^2 + f_7(\alpha,\kappa_1,\kappa_2)AB^4 + f_{8c}(\alpha,\kappa_1,\kappa_2)A \right] \,, \label{resharm}
\end{align}
where
\begin{equation}
        \Delta(\kappa_1,\kappa_2) = 768\kappa_1^3(4\kappa_1^2 - \kappa_2^2)
\label{harmdel}
\end{equation}
and
\begin{align}
        f_1(\alpha,\kappa_1,\kappa_2) &= 3 f_2(\alpha,\kappa_1,\kappa_2)   \,, \notag \\
        f_2(\alpha,\kappa_1,\kappa_2) &= 288\alpha\kappa_1^2(\kappa_2^2 - 4\kappa_1^2) \,, \notag \\
        f_3(\alpha,\kappa_1,\kappa_2,b_1) &= 768\kappa_1^4b_1(-\kappa_2^2 + 4\kappa_1^2) \,, \notag \\
        f_4(\alpha,\kappa_1,\kappa_2) &= 2 f_2(\alpha,\kappa_1,\kappa_2)  \,, \notag \\
        f_5(\alpha,\kappa_1,\kappa_2) &= 45\alpha^2(-\kappa_2^2 + 4\kappa_1^2) \,, \notag \\
        f_6(\alpha,\kappa_1,\kappa_2) &= 2 f_5(\alpha,\delta,\kappa_1,\kappa_2)  \,, \notag \\
        f_7(\alpha,\kappa_1,\kappa_2) &= f_5(\alpha,\delta,\kappa_1,\kappa_2)  \,, \notag \\
        f_{8s}(\alpha,\kappa_1,\kappa_2,b_2) &= f_{non}(\alpha,\kappa_1,\kappa_2) + 32V_1^2(\kappa_2^2 - 4\kappa_1^2)\,, \notag \\ 
        f_{8c}(\alpha,\kappa_1,\kappa_2) &=  f_{non}(\alpha,\kappa_1,\kappa_2) - 160V_1^2(\kappa_2^2 - 4\kappa_1^2)\,, \notag \\ 
        f_{non}(\alpha,\kappa_1,\kappa_2) &= 192\kappa_1^2(V_2^2 - 4\kappa_1^2\kappa_2^2b_2 + 16\kappa_1^4b_2) \,. \label{harmf}
\end{align}
If considering a harmonic resonance with respect to the secondary
lattice wavenumber $\kappa_2$, one obtains the appropriate equations
for the $O(\varepsilon^2)$ slow evolution by switching the roles of
$\kappa_1$ and $\kappa_2$.  Note that the form of equations (\ref{harmf}) corresponds to (\ref{nono}) except for the extra terms in $f_{8c}$ and $f_{8s}$ that arise from the
superlattice.

The equilibria of (\ref{resharm}) are given by Eq.~(\ref{order})
except that one inserts the functions from (\ref{harmf}).
Additionally, the expressions for $C$ and $D$ have $f_{8s}$ rather
than $f_8$ as a prefactor for $B$ and $f_{8c}$ rather than $f_8$ as a
prefactor for $A$.  One also inserts an equilibrium value of $A$ and
$B$ from Eq. (\ref{echo}).  One then inserts equilibrium values of
$A$, $B$, $C$, and $D$ into Eqs.~(\ref{ro}) and (\ref{r1}) to obtain
the spatial profile $R = R_0 + \varepsilon R_1 + O(\varepsilon^2)$ to
use as an initial condition in direct numerical simulations of Eq.
(\ref{nls3}).

\begin{figure}[tbp]
%\centerline{
\epsfig{file=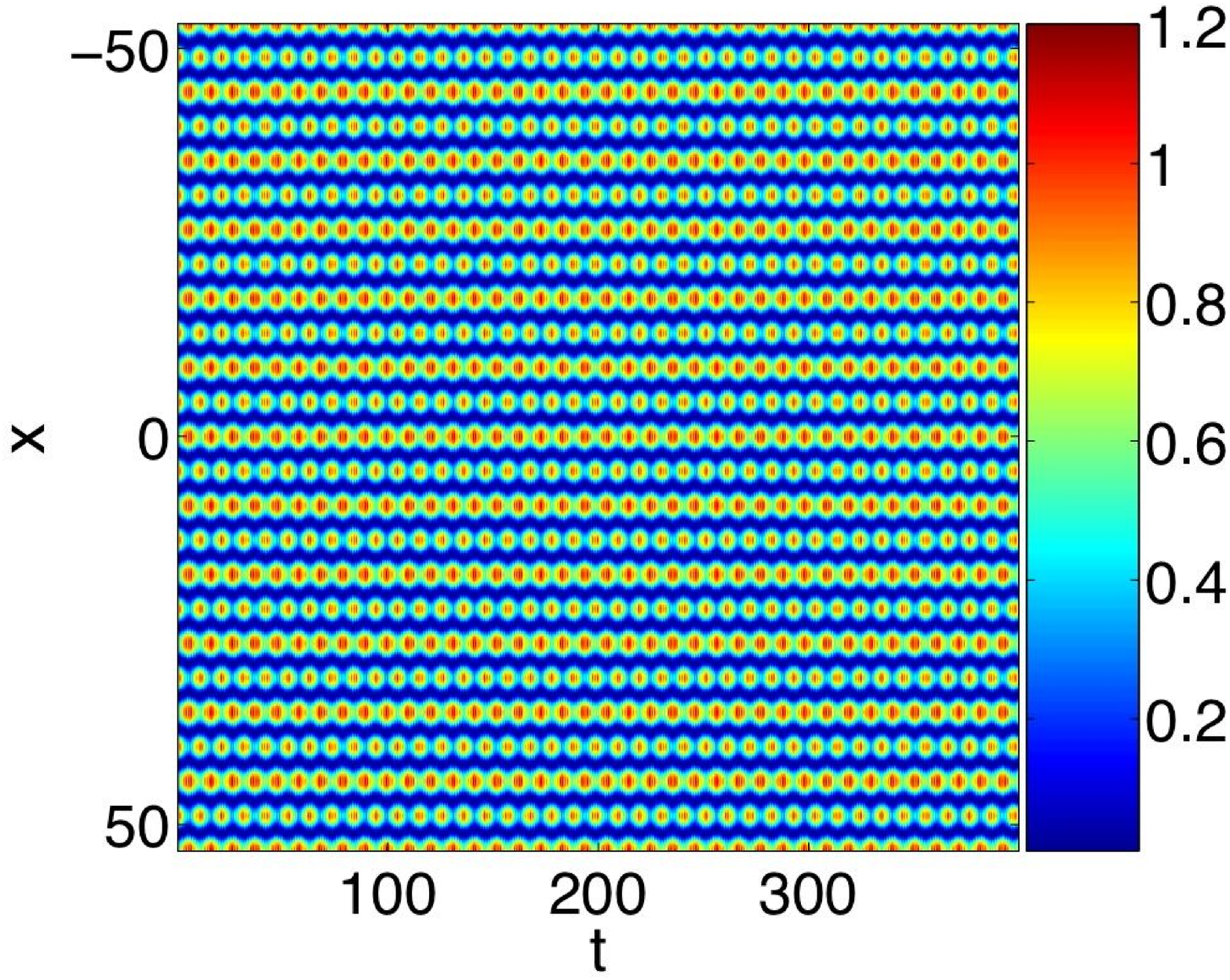, width=7.5cm,angle=0, clip=}
\epsfig{file=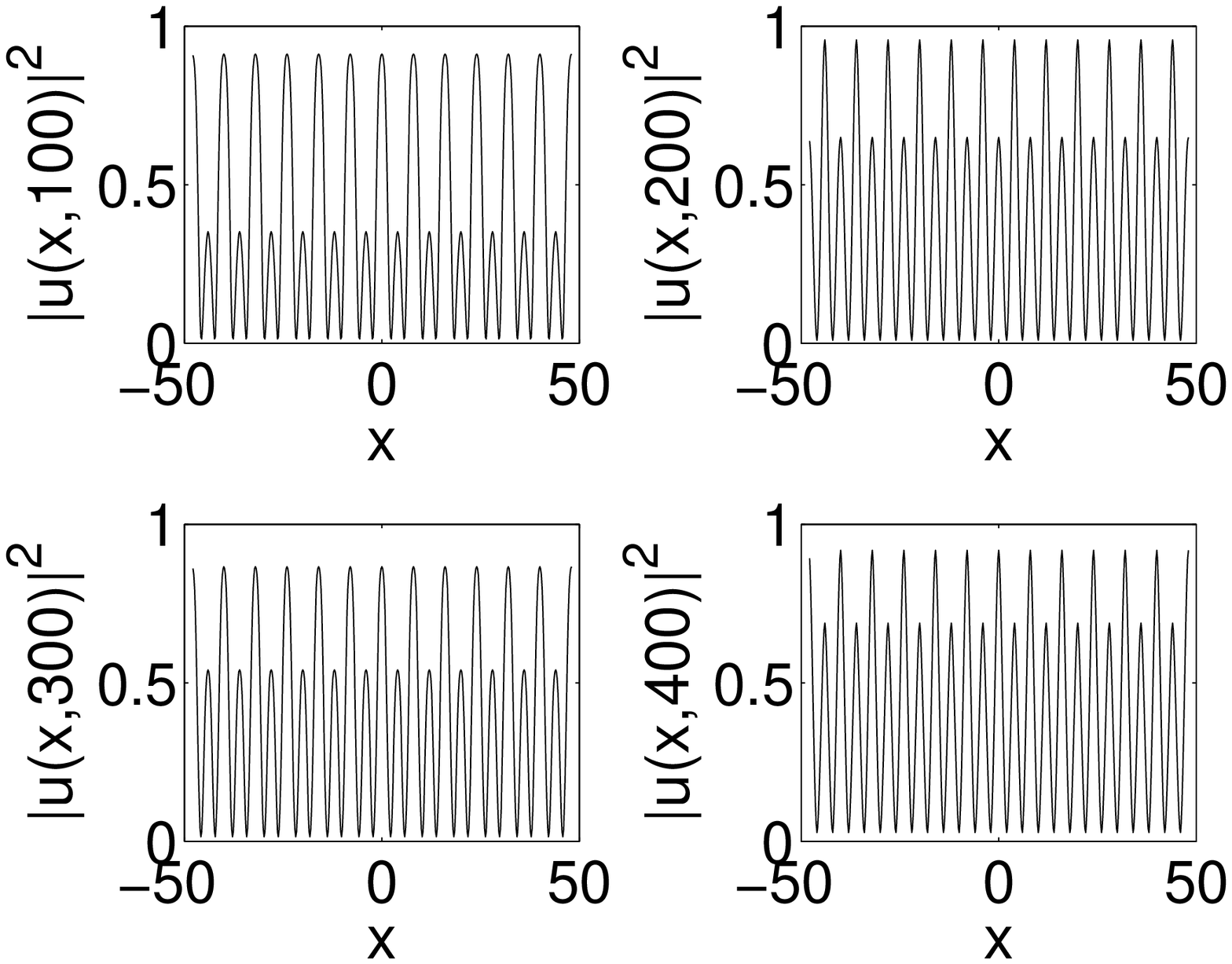, width=7.5cm,angle=0, clip=}
\caption{Same as Fig.~\ref{mpsuper4}, but for the harmonic resonant
  case with respect to the primary lattice wavelength. The solution
  given by Eq.~(\ref{eqr}) is used as an initial condition with $C$
  and $D$ in Eq.~(\ref{r1}) given by Eq.~(\ref{resharm}) with the
  functions (\ref{harmdel},\,\ref{harmf}) [see text for parameter
  details].} \label{mpsuper5}
\end{figure}

A typical example of the single-wavelength resonant case is shown in
Fig.  \ref{mpsuper5}, with $V_2=2 V_1=2$ and $\kappa_2=4 \kappa_1= 4
\sqrt{\delta}= \pi/b$, where $b$ is the stretching factor of Eq.~(\ref{eqb}); we used $b_1=b_2=1$ and $\epsilon=0.1$.  The resulting (spatial) quasiperiodic patterns were generically found to persist
in the dynamics of the system as stable (temporally oscillating)
solutions.

\subsubsection{Ultrasubharmonic Resonances}

Studying BECs in an OSL rather than in a regular OL allows one to
examine the ultrasubharmonic spatial resonances resulting from
interactions between the two lattice wavelengths \cite{nayfeh}. As
with harmonic resonances, an $O(\varepsilon^2)$ calculation is
required to perform the analysis.

When $\kappa_2 + \kappa_1 = \pm 2\sqrt{\delta}$, one has an additive
ultrasubharmonic resonance.  The effective equations governing the
$O(\varepsilon^2)$ slow evolution in this case are (\ref{resharm})
with
\begin{equation}
        \Delta(\kappa_1,\kappa_2) = 32\kappa_1\kappa_2(\kappa_1 + 2\kappa_2)(2\kappa_1 + \kappa_2)(\kappa_1 + \kappa_2)^3 \label{plusdel}
\end{equation}
and
\begin{align}
        f_1(\alpha,\kappa_1,\kappa_2) &= 3 f_2(\alpha,\kappa_1,\kappa_2)   \,, \notag \\
        f_2(\alpha,\kappa_1,\kappa_2) &= -24\alpha\kappa_1\kappa_2[2(\kappa_1^4 + \kappa_2^4) + 9(\kappa_1^3 + \kappa_2^3) + 14\kappa_1^2\kappa_2^2]\,, \notag \\
        f_3(\alpha,\kappa_1,\kappa_2,b_1) &= 16\kappa_1\kappa_2b_1[2(\kappa_1^6 + \kappa_2^6) + 13\kappa_1\kappa_2(\kappa_1^4 + \kappa_2^4) + 34\kappa_1^2\kappa_2^2(\kappa_1^2 + \kappa_2^2) + 46\kappa_1^3\kappa_2^3] \,, \notag \\
        f_4(\alpha,\kappa_1,\kappa_2) &= 2 f_2(\alpha,\kappa_1,\kappa_2)  \,, \notag \\
        f_5(\alpha,\kappa_1,\kappa_2) &= 15\alpha^2\kappa_1\kappa_2[5\kappa_1\kappa_2 + 2(\kappa_1^2 + \kappa_2^2)] \,, \notag \\
        f_6(\alpha,\kappa_1,\kappa_2) &= 2 f_5(\alpha,\delta,\kappa_1,\kappa_2)  \,, \notag \\
        f_7(\alpha,\kappa_1,\kappa_2) &= f_5(\alpha,\delta,\kappa_1,\kappa_2)  \,, \notag \\
        f_{8s}(\alpha,\kappa_1,\kappa_2,b_2) &= f_{non}(\alpha,\kappa_1,\kappa_2) - f_{res}(\alpha,\kappa_1,\kappa_2)\,, \notag \\ 
        f_{8c}(\alpha,\kappa_1,\kappa_2,b_2) &=  f_{non}(\alpha,\kappa_1,\kappa_2) + f_{res}(\alpha,\kappa_1,\kappa_2) \,, \notag \\ 
        f_{non}(\alpha,\kappa_1,\kappa_2) &= 16[13\kappa_1^2\kappa_2^2b_2(\kappa_1^4 + \kappa_2^4) + 46\kappa_1^4\kappa_2^4b_2 + 5\kappa_1^2\kappa_2^2(V_1^2 + V_2^2) \notag \\ &\quad + 2\kappa_1\kappa_2(V_2^2\kappa_1^2 + V_1^2\kappa_2^2+\kappa_1^6b_2 + \kappa_2^6b_2) + 34\kappa_1^3\kappa_2^3b_2(\kappa_1^2 + \kappa_2^2) \notag \\ &\quad + 4\kappa_1\kappa_2(V_1^2\kappa_1^2 + V_2^2\kappa_2^2) + V_1^2\kappa_1^4 + V_2^2\kappa_2^4]\,, \notag \\
        f_{res}(\alpha,\kappa_1,\kappa_2) &= 32V_1V_2[7\kappa_1^2\kappa_2^2 + (\kappa_1^4 + \kappa_2^4) + 4\kappa_1\kappa_2(\kappa_1^2 + \kappa_2^2)]  \,.
\label{plusf}
\end{align}
Note that all the terms in $f_{res}$ are proportional to $V_1V_2$, as they 
arise from the effects of interacting lattice wavelengths.

Equilibria in this situation again satisfy (\ref{order}) except that
one now inserts functions from (\ref{plusdel},\,\ref{plusf}).  Again,
the expressions for $C$ and $D$ have $f_{8s}$ rather than $f_8$ as a
prefactor for $B$ and $f_{8c}$ rather than $f_8$ as a prefactor for
$A$.  One again inserts an equilibrium value of $A$ and $B$ from Eq.
(\ref{echo}).  One then inserts equilibrium values of $A$, $B$, $C$,
and $D$ into Eqs. (\ref{ro}) and (\ref{r1}) to obtain the initial
spatial profile $R = R_0 + \varepsilon R_1 + O(\varepsilon^2)$.
%to utilize as an initial wavefunction in numerical simulations of 
%(\ref{nls3}).

\begin{figure}[tbp]
%\centerline{
\epsfig{file=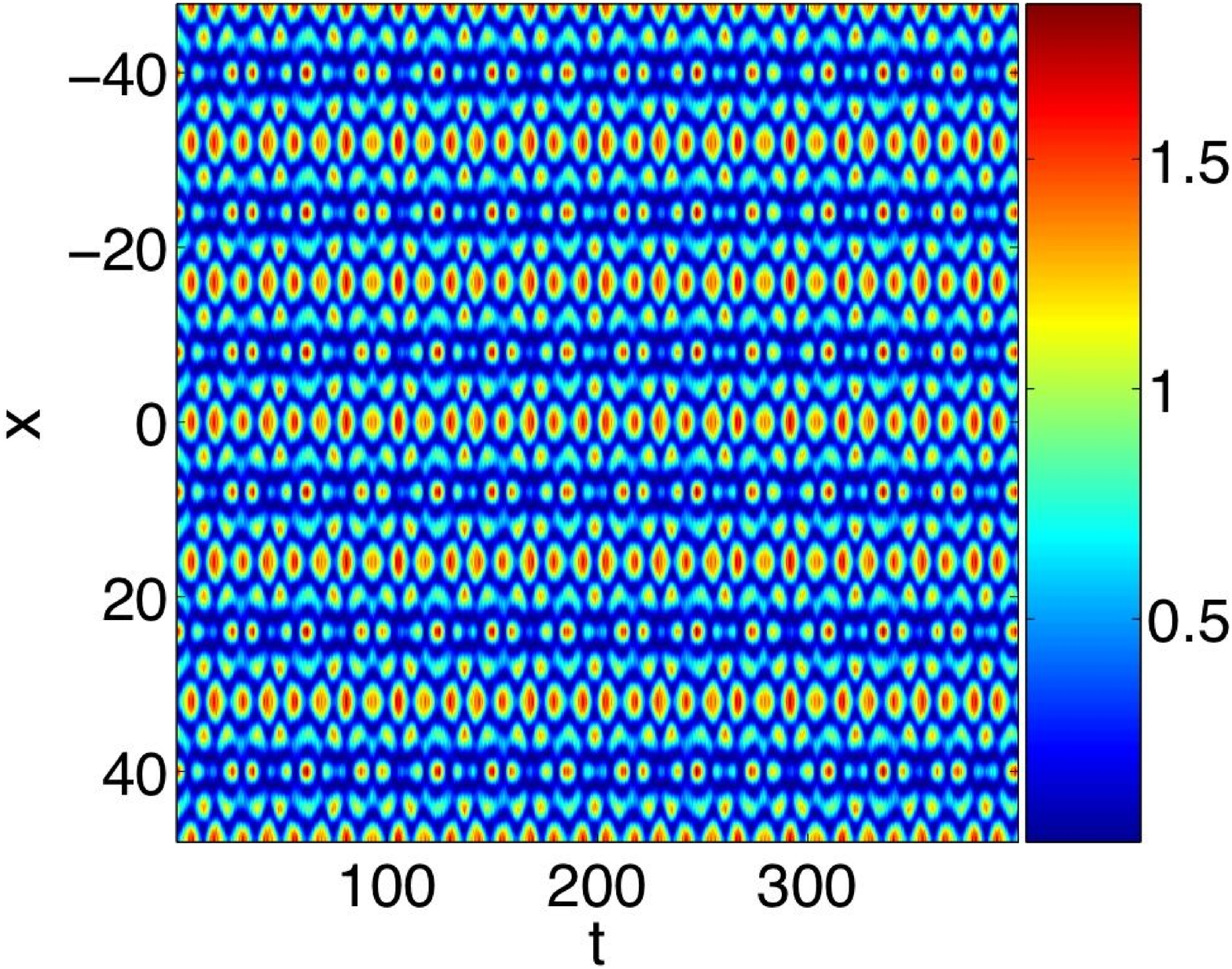, width=7.5cm,angle=0, clip=}
\epsfig{file=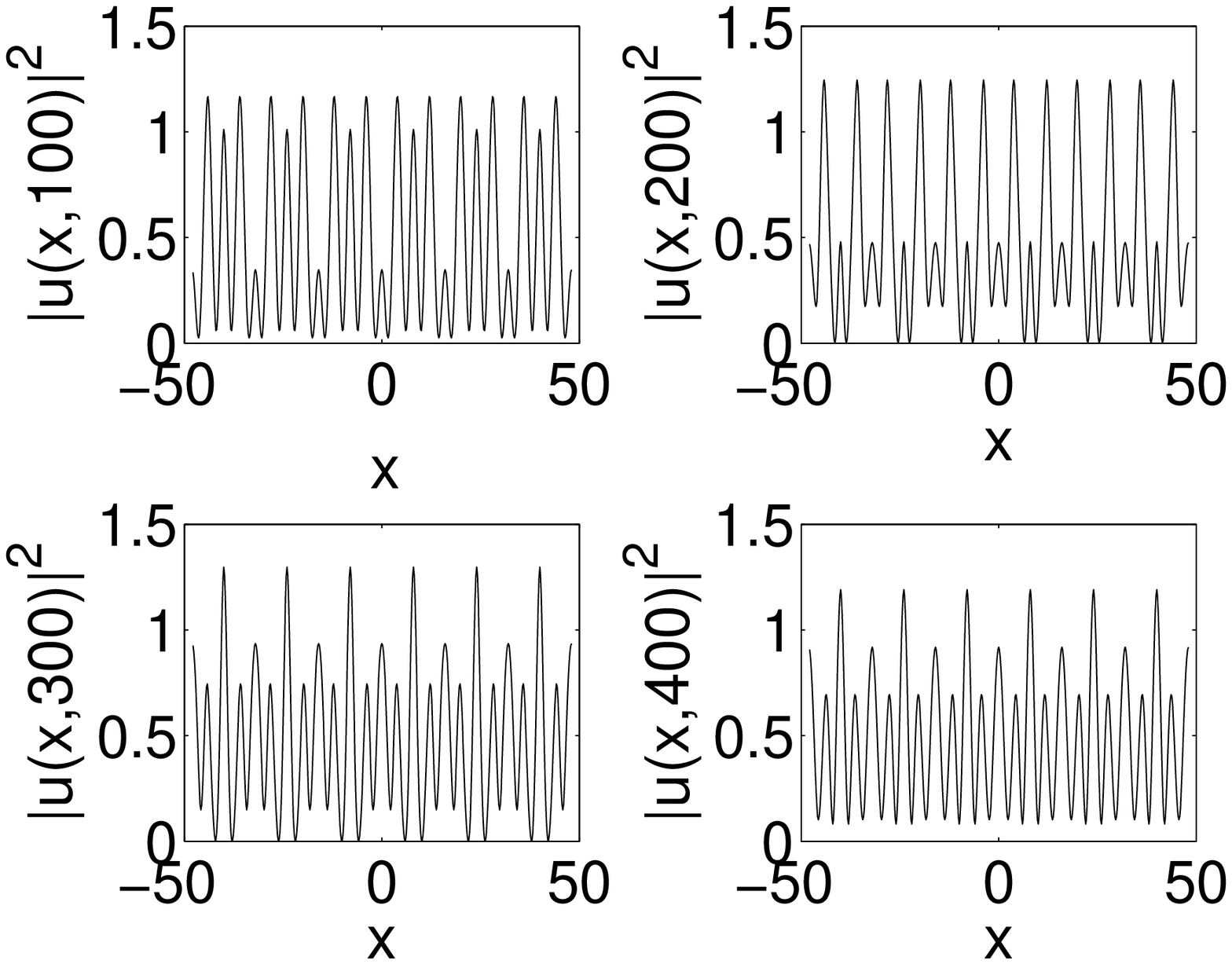, width=7.5cm,angle=0, clip=}
%{\epsfig{file=map7c.ps, width=7cm,angle=0, clip=}}
\caption{Same as Fig.~\ref{mpsuper4}, but for an additive 
  ultrasubharmonic resonance, which arises from the interaction of
  the BEC's two wavelengths. The solution of Eq.~(\ref{eqr}) is used
  as an initial condition with $C$ and $D$ in Eq.~(\ref{r1}) given by Eq.~(\ref{resharm}) with the functions (\ref{plusdel},\,\ref{plusf}) [see text for parameter details].}
\label{mpsuper6}
\end{figure}

A typical simulation of an ultrasubharmonic resonance is shown in
Fig.~\ref{mpsuper6}, with $V_2=2 V_1=2$ and $\kappa_2=3 \kappa_1= 3
\sqrt{\delta}/2= 3 \pi/ (8 b)$, where $b$ is again given by Eq.~(\ref{eqb}) with $b_1=b_2=1$ and $\epsilon=0.1$.  The resulting complex patterns were found to persist as stable dynamical structures
(with periodic time dynamics).

When $\kappa_2 - \kappa_1 = \pm 2\sqrt{\delta}$, one has a subtractive 
ultrasubharmonic resonance.  The effective equations governing the 
$O(\varepsilon^2)$ slow evolution in this case are again (\ref{resharm}), with
\begin{equation}
        \Delta(\kappa_1,\kappa_2) = 32\kappa_1\kappa_2(\kappa_1 - 2\kappa_2)(2\kappa_1 - \kappa_2)(\kappa_1 - \kappa_2)^3  \label{minusdel}
\end{equation}
and
\begin{align}
        f_1(\alpha,\kappa_1,\kappa_2) &= 3 f_2(\alpha,\kappa_1,\kappa_2)   \,, \notag \\
        f_2(\alpha,\kappa_1,\kappa_2) &= 24\alpha\kappa_1\kappa_2[-2(\kappa_1^4 + \kappa_2^4) + 9(\kappa_1^3 + \kappa_2^3) - 14\kappa_1^2\kappa_2^2]\,, \notag \\
        f_3(\alpha,\kappa_1,\kappa_2,b_1) &= 16\kappa_1\kappa_2b_1[2(\kappa_1^6 + \kappa_2^6) - 13\kappa_1\kappa_2(\kappa_1^4 + \kappa_2^4) + 34\kappa_1^2\kappa_2^2(\kappa_1^2 + \kappa_2^2) - 46\kappa_1^3\kappa_2^3] \,, \notag \\
        f_4(\alpha,\kappa_1,\kappa_2) &= 2 f_2(\alpha,\kappa_1,\kappa_2)  \,, \notag \\
        f_5(\alpha,\kappa_1,\kappa_2) &= 15\alpha^2\kappa_1\kappa_2[-5\kappa_1\kappa_2 + 2(\kappa_1^2 + \kappa_2^2)] \,, \notag \\
        f_6(\alpha,\kappa_1,\kappa_2) &= 2 f_5(\alpha,\delta,\kappa_1,\kappa_2)  \,, \notag \\
        f_7(\alpha,\kappa_1,\kappa_2) &= f_5(\alpha,\delta,\kappa_1,\kappa_2)  \,, \notag \\
        f_{8s}(\alpha,\kappa_1,\kappa_2,b_2) &= f_{non}(\alpha,\kappa_1,\kappa_2) - f_{res}(\alpha,\kappa_1,\kappa_2)\,, \notag \\ 
        f_{8c}(\alpha,\kappa_1,\kappa_2,b_2) &=  f_{non}(\alpha,\kappa_1,\kappa_2) + f_{res}(\alpha,\kappa_1,\kappa_2) \,, \notag \\ 
        f_{non}(\alpha,\kappa_1,\kappa_2) &= 16[-13\kappa_1^2\kappa_2^2b_2(\kappa_1^4 + \kappa_2^4) - 46\kappa_1^4\kappa_2^4b_2 - 5\kappa_1^2\kappa_2^2(V_1^2 + V_2^2) \notag \\ &\quad + 2\kappa_1\kappa_2(V_2^2\kappa_1^2 + V_1^2\kappa_2^2+\kappa_1^6b_2 + \kappa_2^6b_2) + 34\kappa_1^3\kappa_2^3b_2(\kappa_1^2 + \kappa_2^2) \notag \\ &\quad + 4\kappa_1\kappa_2(V_1^2\kappa_1^2 + V_2^2\kappa_2^2) - V_1^2\kappa_1^4 - V_2^2\kappa_2^4]\,, \notag \\
        f_{res}(\alpha,\kappa_1,\kappa_2) &= 32V_1V_2[-7\kappa_1^2\kappa_2^2 - (\kappa_1^4 + \kappa_2^4) + 4\kappa_1\kappa_2(\kappa_1^2 + \kappa_2^2)]  \,.
\label{minusf}
\end{align}
As with the additive ultrasubharmonic resonance, all the terms in
$f_{res}$ are proportional to $V_1V_2$.

Equilibria in this case again satisfy (\ref{order}) except that one
inserts the functions from (\ref{minusdel},\,\ref{minusf}).  Recall
once more that the expressions for $C$ and $D$ have $f_{8s}$ rather
than $f_8$ as a prefactor for $B$ and $f_{8c}$ rather than $f_8$ as a
prefactor for $A$.  One also inserts an equilibrium value of $A$ and
$B$ from Eq.~(\ref{echo}).  One then inserts equilibrium values of
$A$, $B$, $C$, and $D$ into Eqs.~(\ref{ro}) and (\ref{r1}) to obtain a
spatial profile $R = R_0 + \varepsilon R_1 + O(\varepsilon^2)$ to
utilize as an initial wavefunction in numerical simulations of
(\ref{nls3}). In this case, the numerical simulations yielded similar
(stable) temporal dynamics as for additive ultrasubharmonic resonances.

\section{Hamiltonian Perturbation Theory and Subharmonic 
Resonances}\label{subby}

In this section, we build on recent work \cite{mapbecprl,mapbec} and
apply Hamiltonian perturbation theory to (\ref{sup}) to
examine period-multiplied wavefunctions and spatial subharmonic
resonances in repulsive BECs loaded into OSL potentials.  (For
expository reasons, we repeat some details of the derivation from those
works in the present one.)  We perturb off elliptic function solutions
of the underlying integrable system and study $2n\!:\!1$ spatial
resonances with a leading-order perturbation method.  Perturbing off
simple harmonic functions, by contrast, requires a perturbative method
of order $n$ to study $2n\!:\!1$ resonances.  At the center of KAM
islands lie `period-multiplied' states.  When $n = 1$, one obtains
period-doubled states in $u$ corresponding to $2\!:\!1$ subharmonic
resonances.  Our analysis reveals period-multiplied solutions of the GP (\ref{nls3}) with
respect to both the primary and secondary sublattice.  

The dynamical systems perspective on period-doubled states and
their generalizations for BECs in OSL potentials given here complements theoretical and experimental work by other authors for the case of regular OL potentials.  In recent experiments, Gemelke {\it et al}. \cite{chu} constructed period-doubled wavefunctions, which have received increased attention (for regular lattices) during the past two years.  In earlier work, Smerzi {\it et al}. \cite{smerzi} reported theoretical studies of spatial period-doubling in the context of modulational (``dynamical'') instabilities of Bloch states and Cataliotti {\it et al}.  \cite{cata} reported experimental observations of superfluid current disruption in chains of weakly coupled BECs.  Period-doubled states, interpreted as soliton
trains, then arise from dynamical instabilities of the energy bands
associated with Bloch states \cite{pethick2}.

\subsection{Unforced Duffing Oscillator} \label{duffing}

We employ exact elliptic function solutions of Duffing's
equation [Eq. (\ref{sup}) with $V_1 = V_2 = 0$], so we no longer
need to assume the coefficient of the nonlinearity is small.
Therefore, we use the ODE
\begin{equation}
        R'' + \delta R + \alpha R^3 
+ \varepsilon R[V_1\cos(\kappa_1 x) + V_2\cos(\kappa_2 x)] = 0 \,, \label{dyn}
\end{equation}
which is just like (\ref{sup}) except that $\alpha$ no longer has the
prefactor $\varepsilon$.

When $\varepsilon = 0$, solutions of (\ref{dyn}) are expressed exactly
in terms of elliptic functions (see, e.g., \cite{zounes,mapbec}
and references therein):
\begin{equation}
        R = \sigma \rho \, \mbox{cn}(u,k) \,, 
\label{ell0}
\end{equation}
where
\begin{align}
        u &= u_1x + u_0 \,, \quad u_1^2 = \delta + \alpha \rho^2 \,, \notag \\
k^2 &= \frac{\alpha \rho^2}{2(\delta + \alpha \rho^2)} \,, \notag \\ 
u_1 &\geq 0, \hspace{.1 in} \rho \geq 0\, , \hspace{.1 in} k^2 \in 
\mathbb{R} \,, \hspace{.1 in} \sigma \in \{-1,1\} \,, \label{ellipse}
\end{align}
and $u_0$ is obtained from an initial condition (and can be set to $0$
without loss of generality).  When $u_1 \in \mathbb{R}$, the solutions
given by (\ref{ellipse}) are periodic.  When $k^2 < 0$, which is the
case for repulsive BECs with positive chemical potentials, equation
(\ref{ellipse}) is interpreted using the reciprocal complementary
modulus transformation (as discussed in Ref.~\cite{mapbec}).

Equation (\ref{dyn}) is integrated when $\varepsilon = 0$ to yield the 
Hamiltonian
\begin{equation}
        \frac{1}{2}R'^2 + \frac{1}{2}\delta R^2 + \frac{1}{4}\alpha R^4 = h\,, 
\end{equation}
with given energy
\begin{equation}
        h = \frac{1}{4}\rho^2(2\delta + \alpha\rho^2) 
= \frac{\delta^2}{\alpha}\frac{k^2 k'^2}{(1-2k^2)^2} \,,
\end{equation}
where $k'^2 := 1 - k^2$.  

The center at $(0,0)$ satisfies $h = \rho^2 = k^2 = 0$.  The saddles
at $(\pm\sqrt{-\delta/\alpha},0)$ and their adjoining separatrix
(consisting of two heteroclinic orbits) satisfy
\begin{equation}
        h = -\frac{\delta^2}{4\alpha} \,, \quad
\rho^2 = \frac{\delta}{|\alpha|} \,, \quad k^2 = -\infty \,.
\end{equation}
The sign $\sigma = +1$ is used for the right saddle and $\sigma = -1$ is 
used for the left one.  Within the separatrix, all orbits are periodic and 
the value of $\sigma$ is immaterial.

\subsection{Action-Angle Variable Description and Transformations}

For the sake of exposition, we construct an action-angle description in steps.  First, we rescale (\ref{dyn}) using the coordinate transformation
\begin{equation}
        \chi = \sqrt{\delta}x \,, \qquad r = \sqrt{-\frac{\alpha}{\delta}}R  
\label{nondim}
\end{equation}
to obtain
\begin{equation}
        r'' + r - r^3 = 0 \label{dyn2}
\end{equation}
when $V_1 = V_2 =  0$.  In terms of the original coordinates,
\begin{equation}
        R(x) = \sqrt{-\frac{\delta}{\alpha}}r\left(\sqrt{\delta}x\right) \,.
\end{equation}

The Hamiltonian corresponding to (\ref{dyn2}) is
\begin{equation}
        H_0(r,s) = \frac{1}{2}s^2 + \frac{1}{2}r^2 - \frac{1}{4}r^4 = h \,, 
\quad h \in [0, 1/4] \,,  \label{sheep}
\end{equation}
where $s := r' = dr/d\chi$.  Additionally, $\rho^2 \in [0,1]$ and
\begin{equation}
        k^2 = \frac{\rho^2}{2(\rho^2 - 1)} \,.
\end{equation}
With the initial condition $r(0) = \rho$, $s(0) = 0$ (which implies
that $u_0 = 0$), solutions to (\ref{dyn2}) are given by
\begin{align}
        r(\chi) &= \rho \, \mbox{cn}\left(\left[1-\rho^2\right]^{1/2}\chi,k\right) \,, \notag \\
        s(\chi) &= -\rho\left[1-\rho^2\right]^{1/2}\,\mbox{sn}\left(\left[1-\rho^2\right]^{1/2}\chi,k\right) \mbox{dn}\left(\left[1-\rho^2\right]^{1/2}\chi,k\right) \,. \label{soly}
\end{align}

The period of a given periodic orbit $\Gamma$ is
\begin{equation}
        T(k) = \oint_\Gamma d\chi = \frac{4K(k)}{\sqrt{1-\rho^2}} \,,
\end{equation}
where $4K(k)$ is the period in $u$ of $\mbox{cn}(u,k)$ \cite{watson}.  The 
frequency of this orbit is
\begin{equation}
        \Omega(k) = \frac{\pi\sqrt{1-\rho^2}}{2K(k)} \,.
\end{equation}

Let $\Gamma_h$ denote the periodic orbit with energy $h = H_0(r,s)$.  The 
area of phase space enclosed by this orbit is constant with respect to 
$\chi$, so we define the action \cite{goldstein}
\begin{equation}
        J := \frac{1}{2\pi}\oint_{\Gamma_h}sdr = \frac{1}{2\pi}\int_0^{T(k)}[s(\chi)]^2d\chi \,,
\end{equation}
which is evaluated to obtain
\begin{equation}
        J = \frac{4\sqrt{1-\rho^2}}{3\pi}\left[E(k) 
- \left(1 - \rho^2/2\right)K(k)\right] \,. 
\end{equation}
The associated angle in the canonical transformation 
$(r,s) \longrightarrow (J,\Phi)$ is
\begin{equation}
        \Phi := \Phi(0) + \Omega(k)\chi \,.
\end{equation}
The frequency $\Omega(k)$ decreases monotonically as $k^2$ goes from 
$-\infty$ to $0$ [that is, as one goes from the separatrix to the center at 
$(r,s) = (0,0)$].  With this transformation, equation (\ref{soly}) becomes
\begin{align}
        r(J,\Phi) &= \rho(J) \, \mbox{cn}\left(2K(k)\Phi/\pi,k\right) 
\,, \notag \\
        s(\chi) &= -\rho(J)\sqrt{1-\rho(J)^2}\,\mbox{sn}\left(2K(k)\Phi/\pi,k\right) \mbox{dn}\left(2K(k)\Phi/\pi,k\right) \,, \label{soly2}
\end{align}
where $k = k(J)$.

After rescaling, the equations of motion for the forced system (\ref{dyn}) take the form
\begin{equation}
        r'' + r - r^3 + \frac{\varepsilon}{\delta}\left[V_1\cos\left(\frac{\kappa_1}{\sqrt{\delta}}\chi\right) + V_2\cos\left(\frac{\kappa_2}{\sqrt{\delta}}\chi\right)\right]r = 0
\end{equation}
with the corresponding Hamiltonian
\begin{align}
        H(r,s,\chi) &= H_0(r,s) + \varepsilon H_1(r,s,\chi) \notag \\
        &= \frac{1}{2}s^2 + \frac{1}{2}r^2 - \frac{1}{4}r^4 + 
\frac{\varepsilon}{2\delta}r^2\left[V_1\cos\left(\frac{\kappa_1}{\sqrt{\delta}}\chi\right) + V_2\cos\left(\frac{\kappa_2}{\sqrt{\delta}}\chi\right)\right] 
\,.
\end{align}
In action-angle coordinates, this becomes
\begin{align}
        H(\Phi,J,\chi) = \frac{1}{2}\rho(J)^2 - \frac{1}{4}\rho(J)^4
        + \frac{\varepsilon}{2\delta}\rho(J)^2 \, \mbox{cn}^2\left(2K(k)\Phi/\pi,k\right)\left[V_1\cos\left(\frac{\kappa_1}{\sqrt{\delta}}\chi\right) + V_2\cos\left(\frac{\kappa_2}{\sqrt{\delta}}\chi\right)\right] \,.  \label{hamaction}
\end{align}

A more convenient action-angle pair $(\phi,j)$ is obtained using the canonical 
transformation $(\Phi,J) \longrightarrow (\phi,j)$, defined by the relations
\begin{equation}
        j(J) = \frac{1}{2}\rho(J)^2 \,, 
\quad \Phi(\phi,j) = \frac{\phi}{J'(j)} \,,
\end{equation}
where
\begin{align}
        k^2 &= \frac{j}{2j-1} \,, \notag \\
        J(j) &= \frac{2}{3}\sqrt{1-2j}\left[\tilde{E}(j) 
- (1-j)\tilde{K}(j)\right] \,, \notag \\
        \tilde{K}(j) &= \frac{2}{\pi}K[k(j)] \,, 
\quad \tilde{E}(j) = \frac{2}{\pi}E[k(j)] \,.
\end{align}
Additionally,
\begin{equation}
        J'(j) := \frac{dJ}{dj} = \sqrt{1-2j}\tilde{K}(j) 
= \frac{1-2j}{\Omega(j)} \,. \label{fre}
\end{equation}
Note that $J \sim j$ for small-amplitude motion.  Furthermore, $j = 0$
at the origin, and $j = 1/2$ on the separatrix.  The Hamiltonian
(\ref{hamaction}) becomes
\begin{align}
        H(\phi,j,\chi) = j - j^2 + \frac{\epsilon}{\delta}j \, \mbox{cn}^2\left(\frac{\tilde{K}(j)}{J'(j)}\phi,k\right)\left[V_1\cos\left(\frac{\kappa_1}{\sqrt{\delta}}\chi\right) + V_2\cos\left(\frac{\kappa_2}{\sqrt{\delta}}\chi\right)\right]\,. \label{haha}
\end{align}

\subsection{Perturbative Analysis}

A subsequent $\mathcal{O}(\epsilon)$ analysis at this stage allows us
to study $2n\!:\!1$ subharmonic resonances for all $n \in \mathbb{Z}$.
Fourier expanding the $\mbox{cn}$ function yields
\begin{equation}
        \mbox{cn}^2\left(\frac{\tilde{K}(j)}{J'(j)}\phi,k\right) 
= \mathcal{B}_0(j) 
+ \sum_{l = 1}^\infty \mathcal{B}_l\cos\left(\frac{2l\phi}{J'(j)}\right) \,,
\end{equation}
where the coefficients $\mathcal{B}_l(j)$ are obtained by 
convolving the Fourier coefficients \cite{zounes,mapbec}, 
\begin{align}
  B_n(j) &= \frac{4}{k(j)\tilde{K}(j)}b_n[k(j)]\,, \notag \\
  b_n(k) &= \frac{1}{2}\mbox{sech}\left[\left(n + 1/2\right)\pi
    K'(k)/K(k)\right] \,, \label{cof1}
\end{align}
of the $\mbox{cn}$ function in Eq. (\ref{haha}), where $K'(k) :=
K(\sqrt{1-k^2})$ is the complementary complete elliptic integral of
the first kind \cite{watson,stegun}.

The resulting $O(\varepsilon)$ term in the Hamiltonian (\ref{haha}) is
\begin{align}
  \varepsilon H_1(\phi,j,\chi) &= \frac{\varepsilon}{\delta}j\mathcal{B}_0(j)\left[V_1\cos\left(\frac{\kappa_1}{\sqrt{\delta}}\chi\right) + V_2\cos\left(\frac{\kappa_2}{\sqrt{\delta}}\chi\right)\right] \notag \\ &\quad + \frac{\varepsilon}{2\delta}jV_1\sum_{l = 1}^\infty \mathcal{B}_l(j)\left[\cos\left(\frac{2l\phi}{J'(j)} + \frac{\kappa_1}{\sqrt{\delta}}\chi\right) + \cos\left(\frac{2l\phi}{J'(j)} - \frac{\kappa_1}{\sqrt{\delta}}\chi\right)\right] \notag \\
&\quad + \frac{\varepsilon}{2\delta}jV_2\sum_{l' = 1}^\infty \mathcal{B}_{l'}(j)\left[\cos\left(\frac{2l'\phi}{J'(j)} + \frac{\kappa_2}{\sqrt{\delta}}\chi\right) + \cos\left(\frac{2l'\phi}{J'(j)} - \frac{\kappa_2}{\sqrt{\delta}}\chi\right)\right] \,. \label{haha2}
\end{align}

The Hamiltonian (\ref{haha2}) is an expansion over infinitely many
subharmonic resonance bands for each of the primary and secondary
lattice wavenumbers.  Each resonance corresponds to a single harmonic
in (\ref{haha2}).  To isolate individual resonances, we apply the
canonical, near-identity transformation \cite{zounes,mapbec}
\begin{align}
        \phi &= Q_i + \varepsilon \frac{\partial W_1}{\partial P} 
+ \mathcal{O}(\varepsilon^2) \,, \notag \\
        j &= P - \varepsilon \frac{\partial W_1}{\partial Q_i} 
+ \mathcal{O}(\varepsilon^2)  \label{lie}
\end{align}
to (\ref{haha2}) with an appropriate generating function $W_1$ that
removes all the resonances except the one of interest.  The
subscript $i$ in $Q_i$ designates whether one is considering a
resonance with respect to the primary or secondary lattice wavenumber.
The transformation (\ref{lie}) is valid in a neighborhood of this
$2n\!:\!1$ resonance and yields an autonomous 1 DOF resonance
Hamiltonian that determines its local dynamics,
\begin{align}
  K(Q,P,\chi;n) &= P - P^2 +
  \frac{\varepsilon}{2\delta}V_iP\mathcal{B}_{n}(P)\cos\left(\frac{2nQ_i}
{J'(P)} - \frac{\kappa_i}{\sqrt{\delta}}\chi\right) +
  \mathcal{O}(\epsilon^2) \,.
\label{kam1}
\end{align}
In focusing on a single resonance band in phase space, one restricts
$P$ to a neighborhood of $P_{n}$, which denotes the location of the
$n$th resonant torus associated with periodic orbits in $2n\!:\!1$
spatial resonance with the primary ($i = 1$) or secondary ($i = 2$)
sublattice (recall that $\kappa_1 < \kappa_2$).

The resonance relation associated with $2n\!:\!1$ resonances with
respect to the $i$th sublattice is \cite{mapbec}
\begin{equation}
        \frac{\kappa_i}{\sqrt{\delta}} = \pm 2n\Omega(P_n)\,. \label{relate}
\end{equation}
Because $\Omega \leq 1$ is a decreasing 
function of $P \in [0,1/2)$, the associated resonance band is present when
\begin{equation}
        \frac{\kappa_i}{\sqrt{\delta}} \leq 2n \,. \label{onset}
\end{equation}
For example, when $\kappa_i = 2.5$ and $\delta = 1$, there are resonances of 
order $4\!:\!1$, $6\!:\!1$, $8\!:\!1$, {\it etc}, but there are no resonances
 or order $2\!:\!1$.  Analytical expressions for the sizes of the resonance bands and the
 locations of their saddles and centers are the same as those obtained
 for BECs loaded into OLs; they are derived in Ref. \cite{mapbec}.  
 
 To examine the time-evolution of period-multiplied solutions, we need
 only the locations of centers, which are obtained by applying one
 more canonical transformation.  We use the generating function
\begin{equation}
        F_i(Q_i,Y,\chi;n) = Q_iY - \frac{\kappa_i}{2n\sqrt{\delta}}J(Y)\chi \,,
\end{equation}
which yields
\begin{align}
  P &= \frac{\partial F_i}{\partial Q_i}(Q_i,Y,\chi) = Y \,, \notag \\
  \xi &= \frac{\partial F_i}{\partial Y}(Q_i,Y,\chi) = Q_i -
  \frac{\kappa_i}{2n\sqrt{\delta}}J'(Y)\chi \,.
\end{align}
The resonance Hamiltonian (\ref{kam1}) becomes
\begin{align}
  K_n(\xi,Y) &= K(Q_i,P,\chi;n)
  + \frac{\partial F_i}{\partial \chi}(Q_i,Y,\chi) \notag \\
  &= Y - Y^2 - \frac{\kappa_i}{2n\sqrt{\delta}}J(Y) +
  \frac{\varepsilon}{2\delta}V_i
  Y\mathcal{B}_n(Y)\cos\left(\frac{2n\xi}{J'(Y)}\right) \,,
  \label{ram}
\end{align}
which is integrable in the $(Y,\xi)$ coordinate system.   

The centers of the KAM islands associated with this resonance occur at
\cite{mapbec}
\begin{equation}
        Y_c = Y_n + \varepsilon \Delta Y + \mathcal{O}(\varepsilon^2)\,, 
\label{ansa}
\end{equation}
where
\begin{equation}
        \Delta Y = \mp\frac{1}{2\delta}\left[\frac{\mathcal{B}_n(Y_n) + Y_n\mathcal{B}_n'(Y_n)}{\Omega(Y_n)\sqrt{1-2Y_n}\tilde{K}'(Y_n) - 1}\right] \,, \label{deldel}
\end{equation}
and the sign is $-$ when $n$ is even and $+$ when $n$ is odd.  One
then converts the value $Y_c$ back to the original coordinates to
obtain an estimate $(R_c,S_c)$ of the location of the center in phase
space.  [One obtains the locations of the other centers associated
with the same resonance band using iterates of $(R_c,S_c)$ under a
Poincar\'e map, but we only need one of these centers for a given
resonance to examine the time-evolution under the GP equation
(\ref{nls3}) of these solutions, which provide the initial
wavefunctions for the PDE simulations.]

\begin{figure}[tbp]
%\centerline{
\epsfig{file=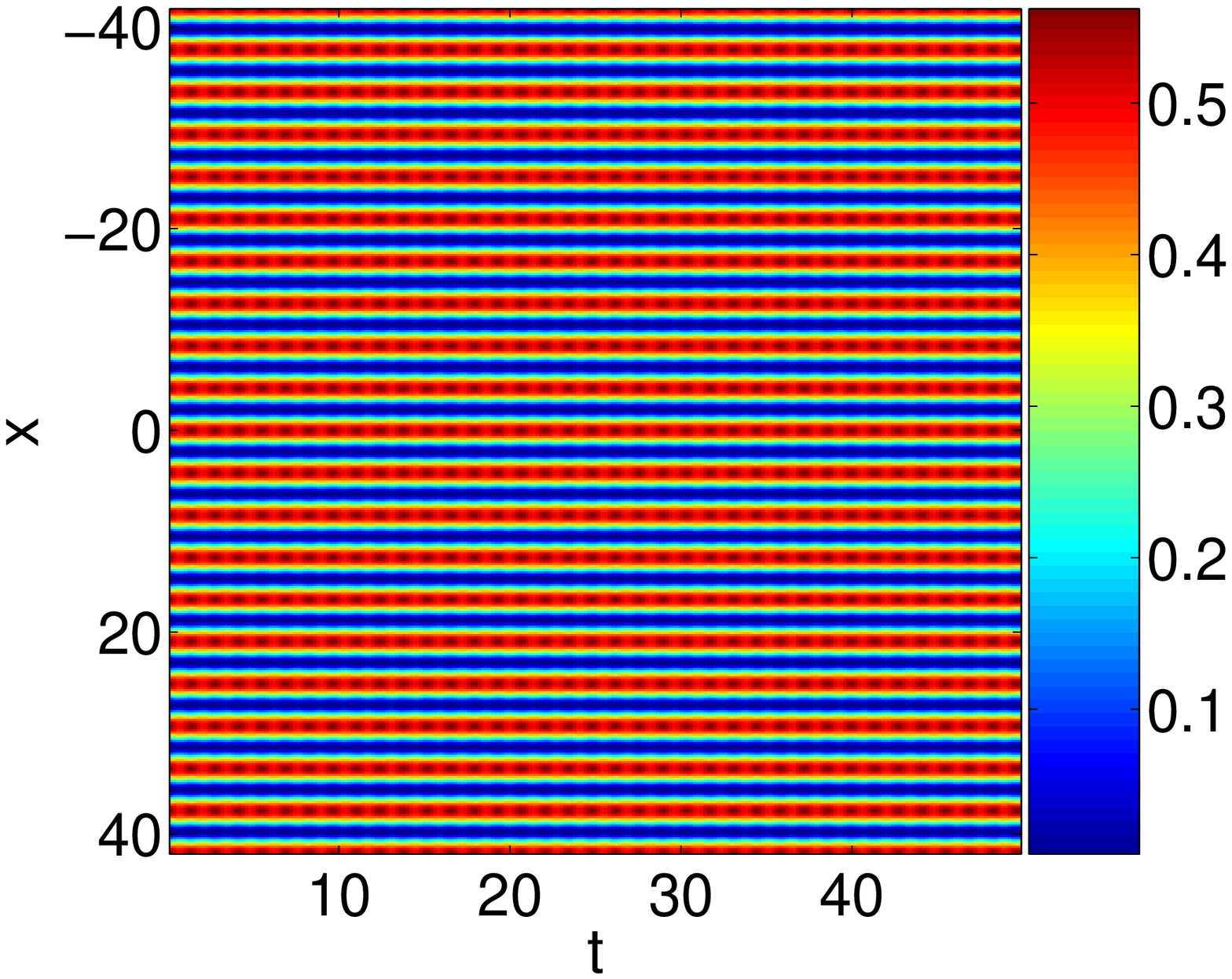, width=7.5cm,angle=0, clip=}
\epsfig{file=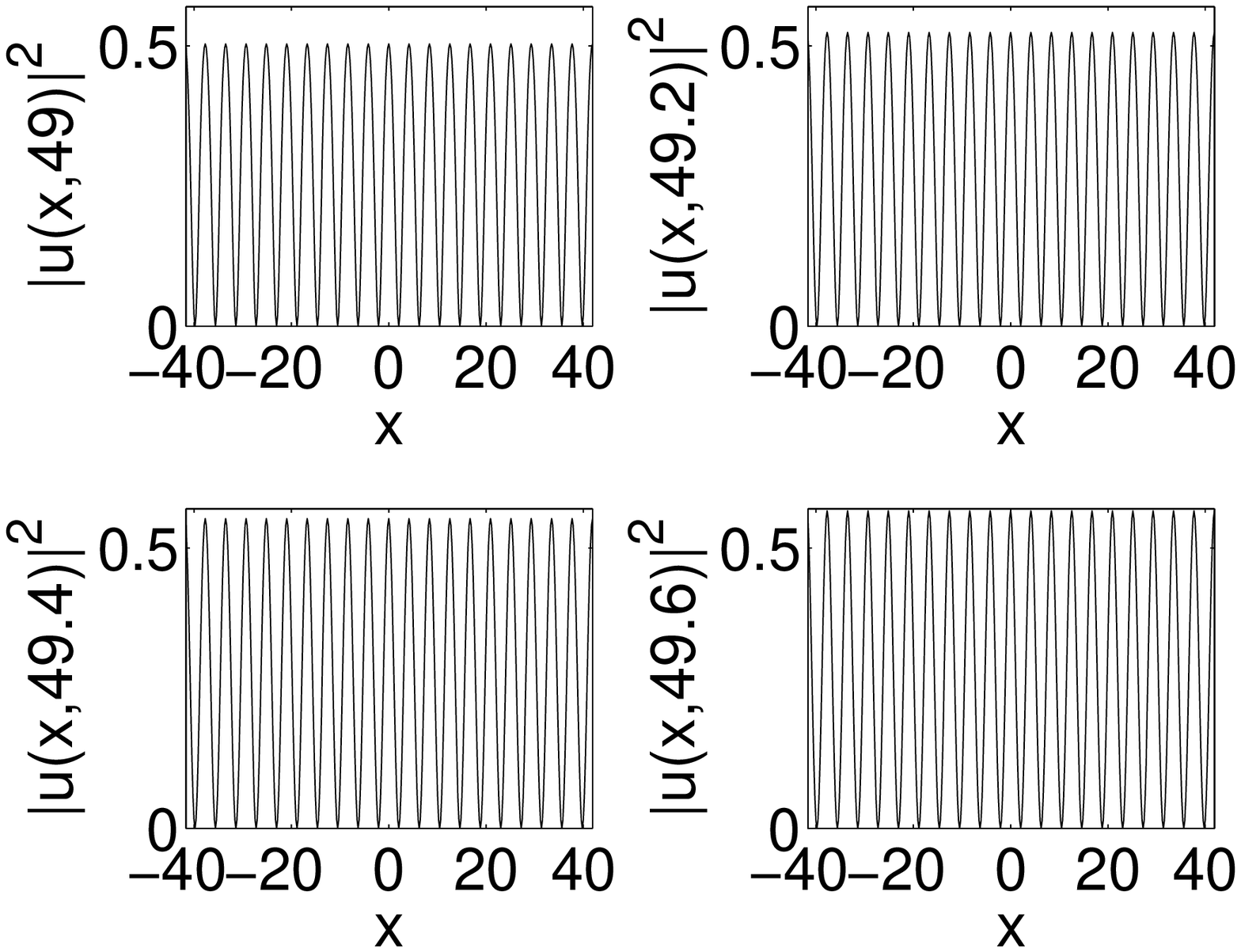, width=7.5cm,angle=0, clip=}
%{\epsfig{file=map7c.ps, width=7cm,angle=0, clip=}}
\caption{Same as Fig.~\ref{mpsuper4}, but for a
  $2\!:\!1$ resonance with respect to the primary sublattice.  The
  solution described in the text [$R = 0.753\cos(\kappa_1 x/2)$ with
  $\kappa_1=1.5=\kappa_2/3$] is used as the initial condition (see the
  text for further parameter details).  The solution appears to be
  dynamically stable and only sustains a small-amplitude oscillation.}
\label{mpsuper7}
\end{figure}

In our numerical computations, we use the parameter values $\hbar = 2m
= 1$, $\delta = 1$, $\alpha = -1$, $\varepsilon = 0.01$, and $V_1 = 1$
in Eqs. (\ref{nls3},\,\ref{dyn}).  With $\kappa = 1.5$, there is a center
for the $2\!:\!1$ resonance with respect to the primary sublattice at
$R_c \approx 0.753$ and $S_c = 0$, so one uses $R = 0.753\cos(\kappa_1
x/2)$ as an initial wavefunction in simulations of (\ref{nls3}) for
any height $V_2$ and wavenumber $\kappa_2$ of the secondary
sublattice. Such a solution is shown in Fig.~\ref{mpsuper7} for
$V_2=2$ and $\kappa_2 = 3\kappa_1$.  It is dynamically stable and
sustains only small amplitude variations (but is otherwise essentially
stationary).  One can similarly examine initial wavefunctions
corresponding to $2\!:\!1$ resonances with respect to the secondary sublattice.

\begin{figure}[tbp]
%\centerline{
\epsfig{file=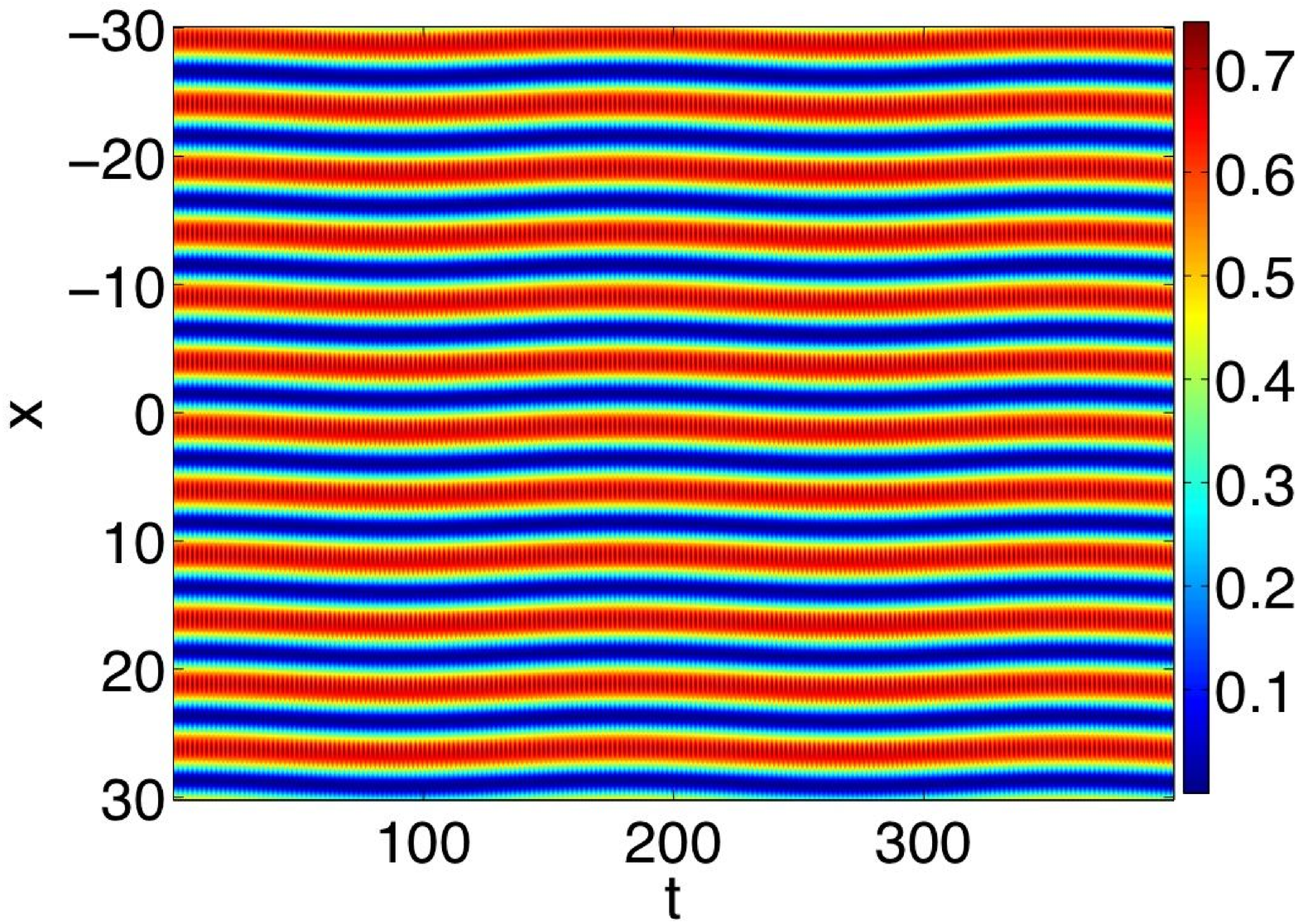, width=7.5cm,angle=0, clip=}
\epsfig{file=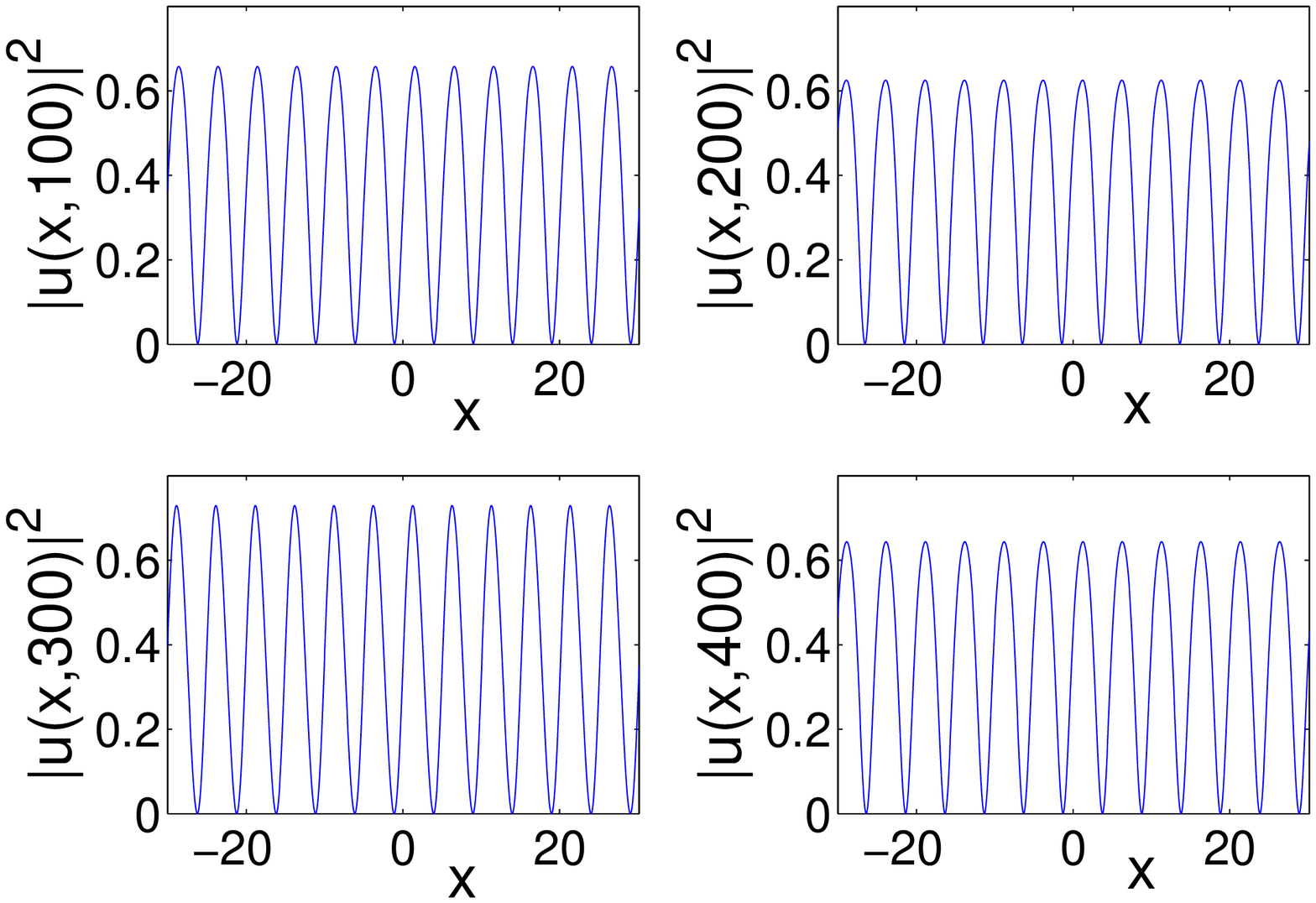, width=7.5cm,angle=0, clip=}
%{\epsfig{file=map7c.ps, width=7cm,angle=0, clip=}}
\caption{Same as Fig.~\ref{mpsuper4}, but for a
  $4\!:\!1$ resonance with respect to the primary sublattice.  The
  solution described in the text [$R = 0.691\cos(\kappa_1 x/4) +
  0.518\sin(\kappa_1 x/4)$ with $\kappa_1=2.5=\kappa_2/3$] is used as
  the initial condition (see the text for further parameter details).
  While structurally stable, the solution pattern appears to be a
  wriggling one, indicating a spatio-temporal breathing.}
\label{mpsuper8}
\end{figure}

With $\kappa_1 = 2.5$, there is a center for the $4\!:\!1$ resonance
with respect to the primary sublattice at $(R_c,S_c) \approx
(0.691,0.324)$, so (recalling the chain rule) one uses $R = 0.691\cos(\kappa_1 x/4) +
0.518\sin(\kappa_1 x/4)$ as an initial wavefunction in simulations of
(\ref{nls3}) The results with $\kappa_2 = 3\kappa_1$ and $V_2 = 2$ are
shown in Fig. \ref{mpsuper8}.  We observe a wriggling pattern in the
contour plot (in the left panel), which indicates (structurally
stable) spatio-temporally oscillatory behavior of the condensate.

With $\kappa_1 = 3.8$, there is a center for the $6\!:\!1$ resonance
with respect to the primary sublattice at $R_c \approx 0.859$ and $S_c
= 0$, so one uses $R = 0.859\cos(\kappa_1 x/6)$ as an initial
wavefunction in simulations of (\ref{nls3}).  We observe that this
period-multiplied state is stable with small-amplitude oscillations,
as was the case for $2\!:\!1$ resonances.  At the same value of
$\kappa_1$, there is a center for the $8\!:\!1$ resonance with respect
to the primary sublattice at $R_c \approx 0.9354$ and $S_c \approx
0.0718$, so one uses $R = 0.9354\cos(\kappa_1 x/8) +
0.151\sin(\kappa_1 x/8)$ as an initial wavefunction.  As was the case
for $4\!:\!1$ resonances, PDE simulations reveal structurally stable
spatio-temporally oscillatory behavior of the condensate (shown for
$4\!:\!1$ resonances as a wriggling pattern in the left panel of
Fig.~\ref{mpsuper8}).  This difference between ``odd'' and ``even''
subharmonic resonances arises from the fact that the former contain
centers on the $R$-axis, whereas the latter do not.  The resulting
initial conditions in the even case hence require both sine and cosine
harmonics, resulting in the observed spatio-temporal breathing.

From a more general standpoint, resonance bands emerge from resonant
KAM tori at action values $P_*$ that satisfy a (three-term) resonance
relation with respect to both sublattices \cite{zasme,qpmatnon},
\begin{equation}
  n_1\frac{\kappa_1}{\sqrt{\delta}} +  n_2\frac{\kappa_2}{\sqrt{\delta}} 
= 2n \Omega(P_*)\,, \label{triple}
\end{equation}
where $n$, $n_1$, and $n_2$ all take integer values.  The
single-sublattice resonance relation (\ref{relate}) is a special case
of (\ref{triple}).

\section{Conclusions}\label{conc}

In this work, we analyzed spatially extended coherent structure
solutions of the Gross-Pitaevskii (GP) equation in optical
superlattices describing the dynamics of cigar-shaped Bose-Einstein
condensates (BECs) in such potentials.  To do this, we derived
amplitude equations governing the evolution of spatially modulated
states of the BEC.  We used second-order multiple scale perturbation
theory to study spatial harmonic resonances with respect to a single
lattice wavenumber, as well as additive and subtractive
ultrasubharmonic resonances.  Harmonic resonances are a second-order
effect that can occur in regular periodic lattices, but
ultrasubharmonic resonances can only occur in superlattice potentials, as they arise from the interaction of multiple lattice wavelengths.  In each situation, we determined the resulting dynamical equilibria,
which represent spatially periodic solutions, and examined the
stability of the corresponding solutions via direct simulations of the
GP equation. In every case considered, the solutions (non-resonant,
resonant with a single wavelength, and resonant due to interactions
between two wavelengths) were found numerically to be dynamically
stable under time-evolution of the GP equation.  Finally, we used
Hamiltonian perturbation theory to construct subharmonically resonant
solutions, whose spatio-temporal dynamics we illustrated numerically
in a number of prototypical cases.

\section*{Acknowledgements}

We wish to acknowledge Todd Kapitula for numerous useful interactions
and discussions during the early stages of this work and the three
anonymous referees and the SIADS editors for several helpful suggestions that improved this
paper immensely.  We also thank Jit Kee Chin and Li You for useful interactions.  P. G. K. gratefully acknowledges support from NSF-DMS-0204585, from the Eppley Foundation for Research, and from an
NSF-CAREER award.  M. A. P. acknowledges support provided by a VIGRE
grant awarded to the School of Mathematics at Georgia Tech, where much of this research was conducted.

%\bibliographystyle{plain}

%\bibliographystyle{siam}
%\bibliography{ref,bec}

\end{document}